\documentclass[useAMS,usenatbib]{mn2e}

\usepackage{times}  % to get nice fonts for PDF
\usepackage{listings}
\usepackage{graphics}
\usepackage{subfigure}
\usepackage{graphicx}
\usepackage{graphicx,xspace}
\usepackage{amssymb}
\usepackage{amsmath}
\usepackage{aas_macros}
\usepackage{lscape}
\usepackage{rotating}
\usepackage{placeins}
\usepackage{natbib}
\usepackage{color}
     
\newcommand{\galform}{{\sc{galform}}\xspace}

\newcommand{\eagle}{{\sc{eagle}}\xspace}
\newcommand{\fire}{{\sc{fire}}\xspace}
\newcommand{\gadget}{{\sc{gadget-3}}\xspace}

\newcommand{\subfind}{{\sc{subfind}}\xspace}

\title[Galactic outflow rates in the EAGLE simulations]
{Galactic outflow rates in the EAGLE simulations}
\author[P. D. Mitchell et al.]{
\newauthor Peter D. Mitchell\thanks{\rm E-mail: mitchell@strw.leidenuniv.nl}$^{1,2}$,
Joop Schaye$^{1}$,
Richard G. Bower$^{3}$,
and Robert A. Crain$^{4}$,
\\
$^{1}$Leiden Observatory, Leiden University, P.O. Box 9513, 2300 RA Leiden, the Netherlands\\
$^{2}$Univ Lyon, Univ Lyon1, Ens de Lyon, CNRS, Centre de Recherche Astrophysique de Lyon UMR5574, F-69230, Saint-Genis-Laval, France\\
$^{3}$Institute for Computational Cosmology, Department of Physics, Durham University, South Road, Durham, DH1 3LE, UK\\
$^{4}$Astrophysics Research Institute, Liverpool John Moores University, 146 Brownlow Hill, Liverpool L3 5RF, UK\\
}
\begin{document}
\date{\today}
\pagerange{\pageref{firstpage}--\pageref{lastpage}} \pubyear{2019}
\maketitle
\label{firstpage}

\begin{abstract}
We present measurements of galactic outflow rates from the \eagle suite of cosmological simulations.
We find that gas is removed from the interstellar medium (ISM) of central galaxies with a dimensionless mass
loading factor that scales approximately with circular velocity as $V_{\mathrm{c}}^{-3/2}$ in
the low-mass regime where stellar feedback dominates. 
Feedback from active galactic nuclei (AGN) causes an upturn in the mass loading for halo masses $> 10^{12} \, \mathrm{M_\odot}$.
We find that more gas outflows through the halo virial radius than is removed from the ISM
of galaxies, particularly at low redshift, implying substantial mass loading within the circum-galactic medium (CGM).
Outflow velocities span a wide range at a given halo mass/redshift, and on average increase positively with redshift
and halo mass up to $M_{200} \sim 10^{12} \, \mathrm{M_\odot}$.
Outflows exhibit a bimodal flow pattern on circum-galactic scales, aligned with the galactic minor axis.
We present a number of like-for-like comparisons to outflow rates from other recent
cosmological hydrodynamical simulations, and show that comparing the propagation of galactic winds as a function of radius
reveals substantial discrepancies between different models.
Relative to some other simulations, \eagle favours a scenario for stellar
feedback where agreement with the galaxy stellar mass function is achieved by removing smaller amounts
of gas from the ISM, but with galactic winds that then propagate and entrain ambient gas out to larger radii.
\end{abstract}

\begin{keywords}
galaxies: formation -- galaxies: evolution -- galaxies: haloes -- galaxies: stellar content
\end{keywords}

\section{Introduction}

In the modern cosmological paradigm, galaxies grow within dark matter haloes, which 
represent collapsed density fluctuations that in turn grow via gravitational instability from
a near-homogeneous initial density field. In this picture, galaxies do not form in 
monolithic formation events, and instead grow gradually via sustained periods of 
gaseous inflow from the larger-scale environment, tracing the hierarchical buildup of
dark matter haloes \cite[e.g.][]{Blumenthal84}. Star formation proceeds
  with sufficient efficiency to deplete gas from the ISM over a timescale that is comparable or shorter
  than a Hubble time, and as such galaxy evolution is to zeroth
  order set by the fluxes of gas into and out of the ISM \cite[e.g.][]{Schaye10,Dave12}.

Observationally, direct measurements of inflowing gas fluxes have remained elusive, with 
only a handful of reported detections \cite[e.g.][]{Rubin12,Fox14,Roberts-Borsani19}. % These references are for down-barrel, Magellanic stream, Stacking galaxy spectra,
Detections and evidence for outflowing gas is comparatively plentiful 
\cite[e.g.][]{Heckman00,Strickland09,Feruglio10,Steidel10,Rubin14,Schroetter16}, % References are for NearUV/Optical (H00,S10,R14), m82 multi-wavelength (S09), molecular AGN (F10)
although determinations of the associated mass flux are likely beset by a number
of systematic uncertainties \cite[e.g.][]{Chisholm16}, and a given outflow tracer probes gas over only a subset
of the relevant spatial scales and gas phases.

The need for substantial outflowing fluxes has long been recognised, for example
in order to explain the form of the observed galaxy luminosity function \cite[e.g.][]{White91,Benson03},
the correlation between galaxy mass and metallicity \cite[e.g.][]{Larson74}, and the presence
of metals in the diffuse intergalactic medium \cite[e.g.][]{Aguirre01}. Feedback
in the form of mass, momentum, and energy input from massive stars and supermassive black holes
is thought to be responsible for driving outflows from galaxies \cite[e.g.][]{Larson74,Silk98}.
These feedback mechanisms are a core element of modern phenomenological models and simulations
that reproduce the observed properties of the overall galaxy population \cite[e.g.][]{Somerville08,Vogelsberger14,Schaye15}.

Determining the efficiency with which galactic winds are driven as a function of the rates at which mass, momentum
and energy are injected into the ISM represents one of the major outstanding challenges of modern astrophysics, both
from the observational and theoretical perspectives. Relevant radiative losses occur in principle
over an enormous dynamic range in scale, and depend on the properties of the ambient medium
over this range. Numerical simulations are routinely used to explore
this problem, again over scales ranging from the small-scale ISM \cite[e.g.][]{Chevalier74,Walch15}, to
the entire galaxy population \cite[][]{Nelson19}, and scales in between \cite[e.g.][]{Hopkins12,Creasey13,Kim18}. % TIGRESS-Kim/Ostriker18

On the large-scale end of this distribution of numerical studies, the \eagle simulation 
project simulates the formation and evolution of galaxies within the full $\Lambda$ Cold 
Dark Matter context, integrating periodic cubic boxes (up to $100^3 \, \mathrm{Mpc^{3}}$ 
in volume) down to $z=0$ \cite[][]{Schaye15,Crain15}. At the reference resolution of the
project, these simulations employ a fiducial baryonic particle mass of $1.81 \times 10^6 \, \mathrm{M_\odot}$, and reach a maximum spatial resolution of 
about $1 \, \mathrm{kpc}$ at $z=0$, and so do not resolve the physics of the ISM.
As with other simulations of this type \cite[e.g.][]{Schaye10,Vogelsberger14,Dubois14,Dave17}, this means
that the \eagle simulations cannot make accurate predictions for the radiative losses that occur
on ISM scales, and a strategy must be adopted to avoid the spurious losses that would occur
should the energy injected by feedback be smoothly distributed. 

In the case of \eagle, spurious losses are mitigated by heating relatively few ISM particles to a high temperature
\cite[$10^{7.5} \, \mathrm{K}$ for stellar feedback, ][]{DallaVecchia12}, with the unresolved radiative losses
then set by hand with model parameters that are calibrated by comparing to various observational 
constraints. As discussed by \cite{Crain15}, it is possible to produce an acceptable
fit to the galaxy stellar mass function inferred from observations by assuming that $100 \, \%$
of the energy available from Type-II supernovae is able to heat gas to high temperatures (in addition
to the energy injection provided by AGN).
To also reproduce the observed distributions
of galaxy sizes as a function of mass, it was found that the energy injected per unit stellar mass had to vary
by factors of a few, scaling negatively with gas metallicity and positively with density.

\eagle is therefore differentiated from a number of similar projects \cite[e.g.][]{Vogelsberger14,Dave17} 
that instead mitigate spurious losses by temporarily decoupling the particles that are kicked by feedback
from the hydrodynamical scheme, while also disabling radiative cooling for these particles. In such alternative
schemes, particles are explicitly kicked with a velocity that scales linearly with the circular velocity 
of the system, and the rate of mass of particles kicked per unit rate of mass of stars formed 
(defining the dimensionless mass loading factor) is assumed to scale negatively with circular velocity.
As no such explicit scaling with galaxy properties is utilised in \eagle\footnote{Beyond the residual dependence of the fraction of energy injected 
on local gas density and metallicity.}, the mass loading and velocities of galactic winds are instead
emergent phenomena, presumably determined (for example) by the escape velocity of system, and the column
density of gas that winds must push through to break out of the ISM. This
  does not guarantee that outflow scalings in the \eagle simulations are necessarily
  more realistic than other schemes used at the same resolution; the \eagle feedback schemes
  are still approximate in nature, and does not explicitly simulate much of the relevant astrophysics that is
  studied analytically or numerically at much higher numerical resolution. Furthermore,
  phenomenological feedback schemes that use decoupled winds only do so temporarily, recoupling
  wind particles once they leave the ISM, meaning that emergent behaviour such as
  anisotropic flow patterns develop despite
  not being imposed at injection \cite[][]{Nelson19}, and the interaction
  between outflows and ambient circum-galactic gas is fully simulated.

We set out in this study to measure the outflow rates of galactic winds from central galaxies in the \eagle simulations.
At a basic level, this allows us to better understand how and why different aspects
of galaxy evolution proceed in a given manner within the simulation, adding valuable
information that can be used to interpret the myriad of other results already published
based on analyses of \eagle. This work also serves as an introduction to a more complete upcoming study 
of the network of inflows, outflows, and recycling of gas flows from \eagle,
and we take care to explain our methodology within this context. For a more
observations-focused analysis of outflows in the \eagle simulations, we refer readers to \cite{Tescari18},
who analyse the simulations within the context of recent integral field unit observations.
In addition, a preliminary version of our inflow rate measurements is used in \cite{Collacchioni19} to
study the connection between inflows and radial metallicity gradients.

On a broader level, we use our measurements of outflow rates to provide a viable quantitative
scenario for how galaxy evolution might proceed across most of the relevant redshift range
and galaxy mass scales. We make the effort to show like-for-like
comparisons with other simulation projects (both large-volume simulations and zoom-in simulations)
to check whether there is yet any consensus emerging from cosmological simulations (the
short answer is that there is little quantitative agreement at present, but there is rough qualitative agreement). All of the
simulations we compare to achieve (to a greater or lesser extent) at least somewhat reasonable 
agreement with the observed stellar properties of galaxies, and so the range of outflow
rates shown in the comparisons might guide observers as well as smaller-scale simulators as to what is likely
required from galactic winds in order to explain the observed galaxy stellar mass function.

The layout of this paper as follows: we introduce our methodology for measuring outflow
rates in Section~\ref{methods_section}, we present measurements of outflow rates
and velocities from \eagle in Section~\ref{results_section}. We finish by placing
our work into the wider context of theoretical models, simulations and observations
in Section~\ref{discussion_section}, and we summarise our results in Section~\ref{summary_section}.

\section{Methods}
\label{methods_section}

\subsection{Rationale}

Our objective is to measure the amount of gas that is ejected from galaxies
and their associated dark matter haloes in the \eagle simulations. This is essential
in order to understand the emergent relationship between stellar mass, gas mass (in
the ISM and also the circum-galactic medium out to the virial radius), and total halo mass.
Outflow rates can be measured from simulations using either Eulerian or Lagrangian methods. The former
involves measuring the instantaneous flux of outflowing gas through a surface (or within a shell)
at a given distance from the center of the galaxy or halo \cite[e.g.][]{DallaVecchia08,Mitchell18_ramses,Nelson19}.
The latter method involves measuring the flux of mass that crosses a surface over a discrete time 
interval \cite[e.g.][]{Neistein12,Christensen16,AnglesAlcazar17}.

We opt to use a Lagrangian method to measure outflow rates. Our primary motivation
for this choice is that the method enables 
accurate measurements of the correct time-integrated
outflow rate of a given galaxy. This is particularly pertinent for the \eagle simulations,
where the high heating temperature used in the subgrid model leads to highly time-variable 
instantaneous outflow rates.
The primary drawback of the Lagrangian method is that correct time-integrated fluxes
are only obtained if fluid elements cross the surface only once over the finite time
interval adopted (fluid elements that cross multiple times cause an underestimate
of the true time-integrated flux). In practice, this means that a substantial number
of simulation outputs (roughly 200 in our case) are required to achieve converged outflow
rates of gas being ejected from the ISM (see Appendix~\ref{ap_converge}), as the timescale
between gas entering and exiting the ISM can be short compared to the halo dynamical time.
As an aside, when we show average radial velocities, or energy and momentum fluxes,
we will switch to Eulerian measurements based on discrete shells; this is because
(unlike mass) these quantities are not necessarily conserved after
leaving the ISM, and so are more clearly defined at a fixed radius.

Another aspect of measuring gas fluxes from simulations is the choice of surface or shell,
and the choice of which subset of the fluid elements flowing through the surface 
should be selected for the measurement.
On the one hand, simple choices for both yield measurements that are easy to reproduce
and compare with other simulations, and the same also applies for comparison with observational 
studies to some extent.
On the other hand, adopting an arbitrary choice of surface runs the risk of not capturing
the desired quantity, which we take to be the flux of gas being removed from the ISM. 
In simulations like \eagle that model the galaxy population across 
a wide range in mass and redshift, the star-forming gaseous content of a galaxy can
vary hugely in structure and spatial scale (both in an absolute sense and relative
to the halo), as is ably demonstrated by the two examples shown in appendix C of
\cite{Mitchell18}. Furthermore, non-negligible amounts of the outflowing flux
on scales close to the ISM can be associated with gas that is moving past pericenter
on orbits that are driven primarily by gravity (rather than by feedback).

For these reasons, we have adopted (and laboriously checked) criteria that select
gas that was within the ISM (at the previous simulation output) and has now (at the current
simulation output) exited the ISM, and is in the process of moving out over a significant distance into the circum-galactic medium.
%These criteria are somewhat more complex compared to measurements at a 
%fixed surface and radial velocity cut \cite[as for example in][]{Nelson19}, but 
%do represent the simplest method we found that still robustly selects the desired 
%outflowing material for all of our simulated galaxy masses and redshifts.
A direct comparison of simple Eulerian measurements with our full Lagrangian
criteria is shown in Appendix~\ref{method_comparison}, for readers who may be interested to see the impact
of our selection criteria on our conclusions.
Our methodology is similar to that of \cite{Christensen16}, who measure gas particles that leave an ISM defined in a similar way using phase
cuts, and that outflow with kinetic energy exceeding that of the gravitational potential,
as well as that of \cite{AnglesAlcazar17}, who perform similar measurements but
instead define the ISM with a Friends-of-Friends algorithm, along with a cut in gas density.

\subsection{Simulations and subgrid physics}

The \eagle project is a suite of hydrodynamical simulations that simulate the formation and evolution of galaxies within the 
context of the $\Lambda$CDM cosmological model \cite[][]{Schaye15}, and that have been publically released \cite[][]{McAlpine16}.
The suite was created using a modified 
version of the \gadget code \cite[last presented in][]{Springel05}, and features a number of cosmological periodic boxes 
containing both gas and dark matter, integrated down to $z=0$. Cosmological parameters are set following \cite{Planck14}, 
with $\Omega_{\mathrm{m}}=0.307$, $\Omega_{\mathrm{\Lambda}}=0.693$, $\Omega_{\mathrm{b}}=0.04825$, $h=0.6777$ and $\sigma_8 = 0.8288$.
The suite employs a state-of-the-art implementation of smoothed particle hydrodynamics
\cite[SPH, see][]{Schaye15,Schaller15a}, and a range of subgrid models which account
for important physical processes that are not resolved by the simulation (radiative cooling, star formation, stellar mass loss and
metal enrichment, supermassive black hole (SMBH) growth, energy injection from stellar and AGN feedback).

Unless otherwise stated, all results presented here are produced using the reference $100^3 \, \mathrm{cMpc^3}$ simulation,
which includes $1504^3$ particles for both gas and dark matter, with particles masses of
$1.81 \times 10^6 \, \mathrm{M_\odot}$ and $9.70 \times 10^7 \, \mathrm{M_\odot}$ for gas and dark matter respectively. This simulation,
\cite[referred to as $L0100N1504$ in][]{Schaye15} uses the subgrid models and parameters of the \eagle reference model
described by \cite{Schaye15} \cite[and also discussed in detail by][]{Crain15}. Hereafter, we refer to this simulation 
as the $100 \, \mathrm{Mpc}$ reference run. In some parts we also utilise smaller $25^3$ and $50^3 \, \mathrm{cMpc^3}$ versions
of the reference simulation (with the same physics and resolution), as well as a $50^3 \, \mathrm{cMpc^3}$ simulation that was simulated without AGN feedback.

An overview of the salient aspects of the \eagle reference model within the context of this study is as follows.
Firstly, stars are allowed to form above the metallicity-dependent threshold for which the gas
is expected to become cold and molecular \cite[][]{Schaye04},

\begin{equation}
n_{\mathrm{H}}^{\star} = \mathrm{min}\left(0.1 \left(\frac{Z}{0.002}\right)^{-0.64},10 \right) \, \mathrm{cm^{-3}},
\label{eqn_sf_thr}
\end{equation}

\noindent where $Z$ is the gas metallicity\footnote{The metal mass fraction, \emph{not} normalised to solar metallicity.}.
Gas particles are artificially pressurized up to a minimum pressure floor set proportional
to gas density as $P \propto \rho_{\mathrm{g}}^{4/3}$, normalized to a temperature of $T = 8 \times 10^3 \, \mathrm{K}$
at a hydrogen density of $n_{\mathrm{H}} = 0.1 \, \mathrm{cm^{-3}}$ \cite[][]{Schaye08}. This acts to ensure that the thermal Jeans mass
is always at least marginally resolved, but prevents the formation of a cold ISM phase. In addition to Eqn~\ref{eqn_sf_thr},
gas particles are eligible to form stars only if they are within $0.5 \, \mathrm{dex}$ in temperature from the temperature floor.

Star formation is implemented stochastically as described in \cite{Schaye08}, with individual gas particles being converted into collisionless star
particles by sampling from a probability distribution such that the star formation rate is given by

\begin{equation}
\psi = m_{\mathrm{gas}} \, A (1 \mathrm{M_\odot} \mathrm{pc}^{-2})^{-n} \, \left(\frac{\gamma}{G} f_{\mathrm{g}} P\right)^{(n-1)/2},
\label{sfr_eqn_eagle}
\end{equation}

\noindent where $m_{\mathrm{gas}}$ is the gas particle mass, $P$ is the local gas pressure,
$\gamma=5/3$ is the ratio of specific heats, $G$ is the gravitational constant, $f_{\mathrm{g}}$ is the gas mass
fraction (set to unity). $A$ and $n$ are taken from the observed Kennicutt-Schmidt star formation law,
$\dot{\Sigma}_\star = A (\Sigma_{\mathrm{g}} / 1 \mathrm{M_\odot pc^{-2}})^n$,
and are set to $A = 1.515 \times 10^{-4} \, \mathrm{M_\odot yr^{-1} kpc^{-2}}$
and $n=1.4$ \cite[][]{Kennicutt98}, with $n$ changed to $n=2$ for hydrogen densities greater than $n_{\mathrm{H}} = 10^3 \, \mathrm{cm^{-3}}$.

Stellar feedback is represented by stochastic thermal energy injection, following the methodology
introduced by \cite{DallaVecchia12}. In this scheme, gas particles are heated by neighbouring
star particles by a fixed temperature jump, $\Delta T = 10^{7.5} \, \mathrm{K}$, with a probability
set such that the average thermal energy injected is $f_{\mathrm{th}} \times \, 8.73 \times 10^{15} \, \mathrm{erg \, g^{-1}}$ of
stellar mass formed, where $f_{\mathrm{th}}$ is a model parameter. For $f_{\mathrm{th}} = 1$,
the injected energy per unit stellar mass corresponds to that of a simple stellar
population with a Chabrier initial mass function (IMF), assuming that $6-100 \, M_\odot$ stars explode as supernovae,
and that each supernova injects $10^{51} \, \mathrm{erg}$ of energy. Neighbouring gas particles
are heated by stellar feedback $30 \, \mathrm{Myr}$ after the formation of a star particle.

In order to empirically recover an adequate match to both the galaxy stellar
mass function and the galaxy size versus stellar mass distribution inferred
from observations \cite[][]{Crain15}, $f_{\mathrm{th}}$ is varied as a function of local 
gas metallicity, $Z$, and the gas density, $n_{\mathrm{H,birth}}$, inherited by the star particle from the gas
from which it formed, with the parametrisation given by

\begin{equation}
f_{\mathrm{th}} = f_{\mathrm{th,min}} + \frac{f_{\mathrm{th,max}} - f_{\mathrm{th,min}}}{1+\left(\frac{Z}{0.1 Z_\odot}\right)^{n_{\mathrm{Z}}} \left(\frac{n_{\mathrm{H,birth}}}{n_{\mathrm{H,0}}}\right)^{-n_{\mathrm{n}}}},
\label{Eagle_eff}
\end{equation}

\noindent where $f_{\mathrm{th,min}}$ and
$f_{\mathrm{th,max}}$ are model parameters that are the asymptotic values
of a sigmoid function in metallicity, with a transition scale at a characteristic
metallicity, $0.1 Z_\odot$ \cite[above which radiative losses are expected to increase due to metal cooling][]{Wiersma09},
and with a width controlled by $n_{\mathrm{Z}}$.
An additional dependence on local gas density is controlled by model parameters,
$n_{\mathrm{H,0}}$, and $n_{\mathrm{n}}$. The two asymptotes, $f_{\mathrm{th,min}}$ and $f_{\mathrm{th,max}}$,
are set to $0.3$ and $3$ respectively, such that between $0.3$ and $3$ times the
canonical supernova energy is injected. $n_{\mathrm{Z}}$ and $n_{\mathrm{n}}$
are both set to $2/\ln(10)$, and $n_{\mathrm{H,0}}$ is set to $0.67 \, \mathrm{cm^{-3}}$.

Supermassive black hole (SMBH) growth is modelled first by seeding SMBH particles at the position of the highest density 
gas particle within dark matter haloes with mass, 
$M_{FOF} > 10^{10} \, \mathrm{M_\odot} /h$, where $M_{FOF}$ is the mass of the friends-of-friends group. 
Black hole particles then accrete
mass with an Eddington limited, Bondi accretion rate that is modified if the accreted
gas is rotating at a velocity which is significant relative to the sound speed \cite[][]{Rosas15,Schaye15}.
Black holes that are sufficiently close to each other in position and velocity are 
allowed to merge, forming a second channel of black hole growth.

Analogous to the implementation of stellar feedback, accreting SMBH particles stochastically
inject thermal energy into neighbouring gas particles \cite[][]{Booth09}, with an energy injection rate

\begin{equation}
\dot{E}_{\mathrm{AGN}} = \epsilon_{\mathrm{f}} \epsilon_{\mathrm{r}} \dot{m}_{\mathrm{acc}} c^2,
\label{Eagle_agn_fb}
\end{equation}

\noindent where $\dot{m}_{\mathrm{acc}}$ is the gas mass accretion rate onto the SMBH,
$c$ is the speed of light, $\epsilon_{\mathrm{r}}$ is the fraction of the
accreted rest mass energy which is radiated (set to $0.1$), and $\epsilon_{\mathrm{f}}$ is a model
parameter which sets the fraction of the radiated energy that couples to
the ISM (set to $0.15$). The injected thermal energy is stored in the SMBH particle 
until it is sufficiently large to, on average, heat a single neighbouring
gas particle by $\Delta T = 10^{8.5} \, \mathrm{K}$, a temperature jump which 
is an order of magnitude larger than the value used for stellar feedback ($\Delta T = 10^{7.5} \, \mathrm{K}$).

\subsection{Relating phenomenological feedback modelling to the underlying astrophysics}

Having described the salient features of the phenomenological star formation and feedback modelling
  used in the \eagle simulations, we briefly discuss here how this relates to the underlying
  astrophysics of feedback. As presented in \cite{Schaye15}, the conceptual intent for the stellar feedback
  model in \eagle is that it represents the combined effects of all stellar feedback processes
  that are thought to be relevant for galaxy evolution, including supernovae, radiative feedback,
  \cite[e.g.][]{Krumholz12,Rosdahl15}, stellar winds \cite[e.g.][]{Gatto17},
  and the effects of cosmic rays seeded by supernovae \cite[e.g.][]{Uhlig12,Booth13,Girichidis16}.
  By imposing a star formation law that reproduces the observed Kennicutt-Schmidt relation, feedback
  is not required to set the local efficiency of star formation, reducing the need for ``early''
  stellar feedback that pre-processes the ISM before SNe explode.
  Furthermore, due to the coarse numerical resolution of \eagle (compared to the afore-mentioned numerical
  studies), and with the equation of state that artificially pressurizes the ISM, the gas phase
  distribution in the ISM is not expected to be realistic (regardless of whether early stellar feedback
  is included), which precludes the robust application
  of much higher resolution calculations that predict (for example) the momentum and energy injection
  for isolated supernovae as a function of local density and metallicity
  \cite[e.g.][]{Cioffi88,Kim15,Walch15,Gentry17,Gentry20}.

Similarly, the AGN feedback model used in \eagle is also a heavily coarse-grained
  description of the underlying astrophysics. The exact mechanism by which energy is coupled
  to the surrounding gas is not specified, and the scheme may also need to mimic the outcome
  of plasma physics in relation to AGN feedback in the intra-cluster medium, such as
  the effects of cosmic rays in heating and providing pressure support
  \cite[e.g.][]{Loewenstein91,Sijacki08,Ruszkowski17}.
  These limitations are important to keep in mind when interpreting results from simulations
  like \eagle. The tradeoff however is that by using simple phenomenological feedback schemes that
  mitigate immediate radiative losses, we can calibrate a simulation to produce a realistic and
  representative population of galaxies across a significant dynamic range in stellar mass, 
  and so present a physically viable scenario for how mass and
  energy fluxes at different scales regulate the growth of galaxies.

% I was considering writing something here about CGM issues - but I think this is best kept for the convergence section
%\textbf{The phenomenological nature of the \eagle feedback model should also be kept
%  in mind when considering the interaction between outflowing winds and the
%  circum-galactic medium (CGM, which we show later plays an important role in the
%  regulation of galactic gaseous assembly in \eagle). As has long been discussed
%  in the literature, the Lagrangian nature of particle codes (and of the
%  usual refinement strategy used in grid codes) means that cosmological simulations
%  are not well suited to resolve the anticipated multi-phase phase distribution
%  of the CGM, with implications for the distribution of cool absorbing gas
%  traced in observations \cite[non-Lagrangian refinement strategies have
%  been used to explore this topic over the last year:][]{Hummels19,Peeples19,Suresh19,VanDeVoort19}.
%  This potentially has implications.
%  As these interactions are also relevant for regulating galaxy assembly,
%  the phenomenological approach of the \eagle feedback model also applies,
%  in the sense that

\subsection{Subhalo identification \& merger trees}

Haloes are first identified from a given simulation output as groups, using a friends-of-friends
(FoF) algorithm, with a dimensionless linking length of $b=0.2$ \cite[][]{Davis85}. FoF groups
are then split into subhaloes using the \subfind algorithm \cite[][]{Springel01,Dolag09}. Each
subhalo consists of a set of bound particles (including gas, stars, black holes and dark matter).
For each FoF group, the subhalo containing the particle with the lowest value of the gravitational
potential is defined as the central subhalo (and galaxy). Other subhaloes within the FoF group
are defined as satellites. The subhalo (and associated galaxy) center is defined as the position
of the particle with the lowest value of the gravitational potential.
Finally, for central subhaloes we take an additional step and add/remove particles that are within/outside
$R_{200}$\footnote{In practice this acts to add gas particles within the virial radius that
have been raised by feedback to sufficiently high internal plus kinetic energy that they are no longer
considered bound to the subhalo by \subfind. We need to keep these particles associated to the subhalo
in order to ensure that our measurements of halo outflow rates are correct.}, provided the 
particles are not associated with another subhalo or FoF group.
Here, $R_{200}$ is the radius enclosing a mean spherical overdensity which is $200$ times the critical density
of the Universe at a given epoch. Halo masses and virial radii quoted throughout this paper are defined
as $M_{200}$ and $R_{200}$ respectively, where $M_{200}$ is the mass enclosed within $R_{200}$.

We construct merger trees using the algorithm described in appendix A of \cite{Jiang14}.
In brief, for each subhalo in a given simulation output (the progenitor in question), the algorithm attempts to identify
a single descendant subhalo in the next simulation output. The descendant is selected
as the subhalo containing the largest fraction of a set of the progenitor's most-bound particles. 
Furthermore, if the largest fraction of a set of the most-bound particles of the descendant
come from the progenitor in question, the progenitor is identified as the main progenitor
of the descendant.
In cases where the progenitor in question is not identified as a main progenitor, a number of later
simulation outputs are also searched in an attempt to find a descendant for which the progenitor
in question is the main progenitor. This procedure accounts for cases where subhaloes temporarily
cannot be identified by \subfind against the backdrop of a larger subhalo.
In post-processing we identify rare cases where the identified main progenitor of a descendant
is a clump identified as a subhalo by \subfind, but is dominated by star and black hole particles, rather
than dark matter particles. In these cases, we find the most massive progenitor of the descendant and set that subhalo as the main
progenitor. Put together, this is then the definition of the main progenitor which we use
throughout our analysis (in the sense that we measure particles that were present in the ISM/halo
of the main progenitor that have since been ejected from the descendant).

We use a number of sets of merger trees constructed with differing numbers of simulation outputs.
Most of our results use trees constructed with $200$ simulation \emph{snipshots}, where
snipshots are simulation outputs that contain a subset of the information available for each
particle from the more sparsely sampled simulation \emph{snapshots}. The temporal spacing
between these $200$ snipshots is shown in Appendix~\ref{ap_converge}.
In some cases, we use merger trees constructed with different numbers of snipshots or snapshots,
either to test the temporal convergence of our method, because processed \subfind outputs
were not available for a given simulation, or because we required particle information that
is only present within the snapshots.

\subsection{Particle partitioning}
\label{partition_subsection}

Within a given subhalo, we partition the baryonic particles into a discrete number of groups.
Firstly, star and black hole particles form two distinct groups. For gas particles, we
select particles belonging to the ISM, with the remainder forming a circum-galactic halo component.

Our ISM selection criteria are closely related to the star formation criteria used in the
simulation. We define the ISM as the sum of:

\begin{itemize}
\item Star-forming gas (i.e. particles with $n_{\mathrm{H}} > n_{\mathrm{H}}^{\star}$ and are within $0.5 \, \mathrm{dex}$ of the temperature floor), irrespective of radius.
\item Gas within $0.5 \, \mathrm{dex}$ of the temperature floor, $T_{\mathrm{EOS}}$\footnote{
  This is the density-dependent temperature floor corresponding to the equation of state imposed onto the ISM: $P_{\mathrm{eos}} \propto \rho^{4/3}$.}
  ($\log_{10}(T) < \log_{10}(T_{\mathrm{EOS}}(\rho_{\mathrm{g}})) +0.5$),
  with density, $n_{\mathrm{H}} > 0.01 \, \mathrm{cm^{-3}}$, and radius, $r/R_{\mathrm{Vir}} < 0.2$.
\end{itemize}

The choice to include non-star-forming gas down to $n_{\mathrm{H}} = 0.01 \, \mathrm{cm^{-3}}$
is made primarily to account for dense gas in low-mass haloes with low metallicity, and in effect
approximately selects neutral hydrogen out to the imposed radius cut \cite[][]{Rahmati13}. The effect of this inclusion
for our results is to significantly enhance the outflow rates of low-mass galaxies (see Appendix~\ref{vcut_sec}), where little
star formation and chemical enrichment has occurred. The inclusion also increases the
specific angular momentum of the ISM (by effectively selecting more diffuse neutral material in the
outskirts of galaxy disks), which we plan to study in the context of
inflows/outflows in future work \cite[see also][]{Mitchell18}. 

We impose a radial cut
for the non-star-forming ISM component to exclude dense and low-metallicity infalling and filamentary circum-galactic material (found
mostly at high redshift). We do not impose any radial cut for star-forming gas in order to account
for stellar feedback that occurs outside of this radius, which is relevant for removing gas
from the star-forming gas reservoir of galaxies at high redshift in the simulation ($z \gtrapprox 2$).

\subsection{Measuring outflow rates}
\label{outflow_description}

%\begin{figure*}
%\begin{center}
%\includegraphics[width=40pc]{Figures/sfh_illustration_figure.pdf}
%\caption{
%Example formation history of the main progenitor of a single galaxy with final stellar mass, $M_\star(z=0) = 1.7 \times 10^10 \, \mathrm{M_\odot}$, plotted as a function of cosmic time. 
%The SFR of the galaxy is shown in black, and the mass outflow rates from the ISM and halo are shown in blue and red respectively. 
%The SMBH mass accretion rate, rescaled by $\epsilon_{\mathrm{bh}} / \epsilon_{\mathrm{sn}}$ (see Eqn.~\ref{??}), is shown in green. This galaxy is taken from the $25 \, \mathrm{Mpc}$ reference run.}
%\label{sfh_illustration}
%\end{center}
%\end{figure*}

%Fig.~\ref{sfh_illustration}. This figures shows an example of mass outflow rates and star formation/black hole accretion rates for a single galaxy.

% Does a discussion of the pros and cons of this method belong here on in the intro - decide later
%Our methodology is based on that of \cite{Neistein12}, but has been updated to pay
%much greater care to selecting particles that are genuinely outflowing.

We use a Lagrangian particle tracking method to measure gas outflow rates from galaxies and haloes.
We define galaxy-scale outflow rates as the summed mass of particles leaving the ISM per unit time,
measured over some finite time interval between two simulation outputs. Halo-scale outflow
rates are then defined accordingly for particles leaving the halo virial radius per unit time.
In both cases, we apply the additional selection criteria described below to check that the particles are genuinely outflowing.
Further details of the rationale, exploration and testing that was used to arrive
at these criteria are described in Appendix~\ref{ap_method}, along with a comparison to simple
shell-based outflow rate measurements.

For both galaxy-scale and halo-scale outflows, we require that outflowing particles satisfy

\begin{equation}
\frac{\Delta r_{21}}{\Delta t_{21}} > 0.25 \, V_{\mathrm{max}},
\label{ism_wind_criteria1}
\end{equation}

\noindent and for galaxy-scale outflows, we also require that

\begin{equation}
v_{\mathrm{rad,}1} > 0.125 \, V_{\mathrm{max}},
\label{ism_wind_criteria2}
\end{equation}

\noindent where $V_{\mathrm{max}}$ is the maximum of circular velocity profile of the halo,
$v_{\mathrm{rad,}1}$ is the instantaneous radial velocity of the particle at the first
simulation output after the particle has left the ISM (output 1).
$\frac{\Delta r_{21}}{\Delta t_{21}}$ is the time-averaged radial velocity, measured
by comparing the particle radius at this output with its radius at a later
simulation output (output 2). We choose the time spacing between outputs 1 and 2 to 
correspond as closely as possible to one quarter of a halo dynamical time\footnote{For simplicity we
approximate the halo dynamical time as $10\%$ of the age of the Universe.}.
This ensures that our selection criteria are capable of achieving converged answers with respect
to the chosen temporal spacing of simulation outputs (see Appendix~\ref{ap_converge}).
Further to Eqns~\ref{ism_wind_criteria1} and \ref{ism_wind_criteria2}, we also select
outflowing particles that have an instantaneous radial velocity greater than $V_{\mathrm{max}}$ (at output 1). This catches (rare) cases where particles are
feedback-accelerated briefly to very high radial velocities but stall\footnote{Such particles
rapidly decelerate due to encountering a dense structure.} before moving a significant
distance out into the halo.

Eqn~\ref{ism_wind_criteria1} is our main criterion for selecting galaxy-scale outflows.
It effectively demands that the particles will move outwards by at least one sixteenth of
the virial radius within one quarter of a halo dynamical time. Eqn~\ref{ism_wind_criteria2} is a less
stringent secondary criterion that helps to ensure that the particle has already
joined the outflow by output 1 (from inspection of particle trajectories we find that 
this is only relevant for galaxy-scale outflows).

Particles that leave the ISM/halo that are not selected as outflowing by the aforementioned
criteria are added to a list of candidate wind particles that are then propagated down
the halo merger tree on subsequent simulation outputs. These particles are re-tested
against the same selection criteria at each subsequent simulation output until they either 
satisfy the criteria or three halo dynamical times have expired (at which point they are removed
from the candidate wind list). This procedure ensures that particles that fluctuate over
the ISM or virial radius boundary are accounted for in the outflow rate measurements should
they be significantly accelerated while just outside the boundary. Including these particles
has a negligible effect on outflow rates for lower mass galaxies ($M_{200} < 10^{12} \, \mathrm{M_\odot}$), 
but does increase the outflow rates of high-mass galaxies appreciably, and becomes the main contribution
to galaxy-scale outflows for halo masses of $M_{200} > 10^{13} \, \mathrm{M_\odot}$.

Our results are not highly sensitive to the exact values adopted for these selection
criteria (as demonstrated in Appendix~\ref{vcut_sec}), although it is important
to include some cut on time-averaged radial velocity.

% Rmax histograms, individual examples - checking purity / copleteness/ convergence
% May want to state the time resolution of the trees here?

\section{Results}
\label{results_section}

\begin{figure*}
\begin{center}
\includegraphics[width=40pc]{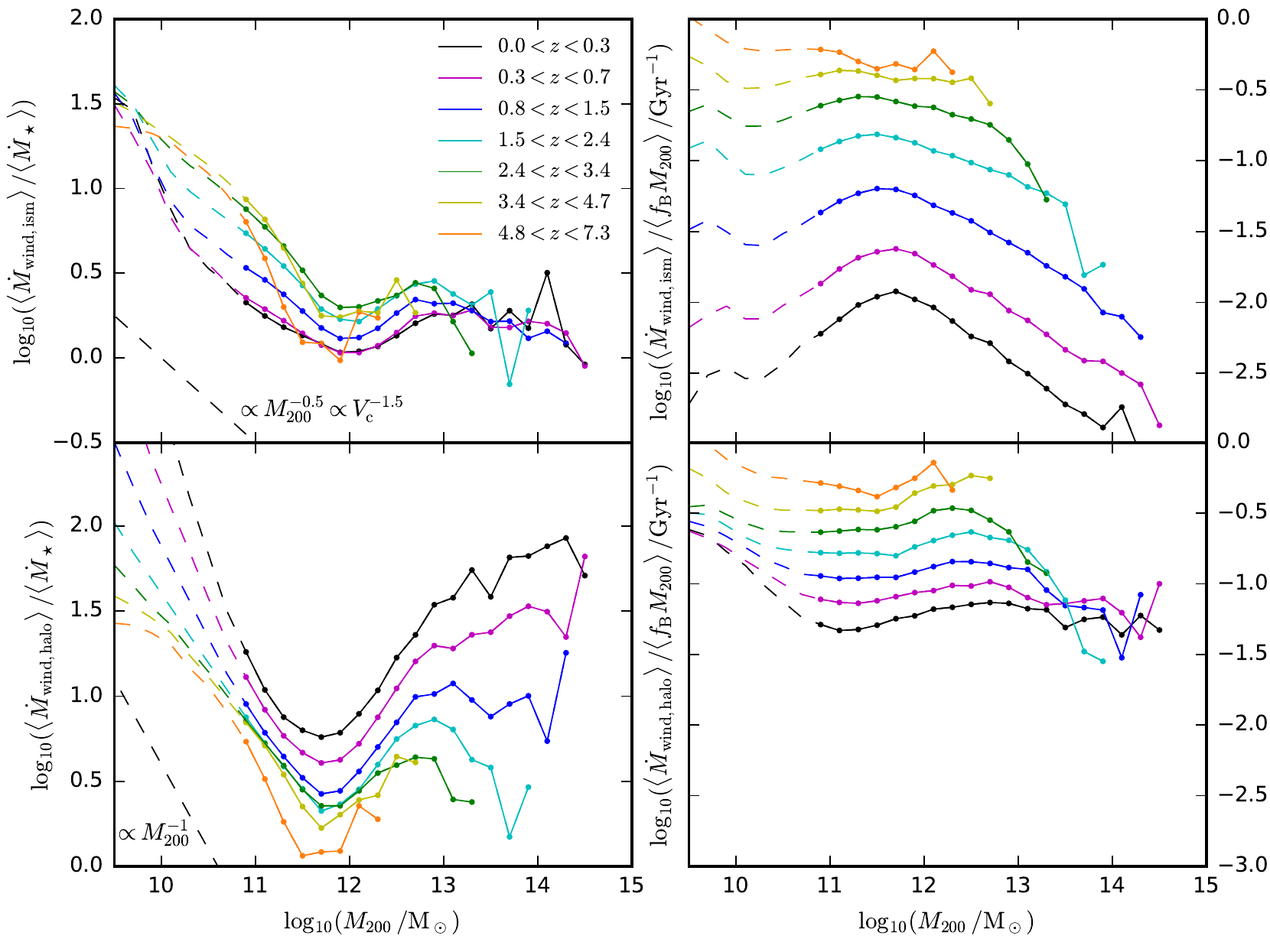}
\caption{Mean mass outflow rates from the ISM (top panels) and haloes (bottom panels) for central galaxies, plotted as a function of halo mass.
Outflow rates are quantified as a dimensionless mass loading factor (mean outflow rate over mean star formation rate, left panels), and as a mean outflow rate per unit halo mass, 
scaled by the cosmic baryon fraction, $f_{\mathrm{B}} \equiv \Omega_{\mathrm{b}} / \Omega_{\mathrm{m}}$ (right panels).
Different line colours correspond to different redshift intervals, as labelled, and mean fluxes and star formation rates are computed across all galaxies in each redshift/mass bin.
Solid (dashed) lines indicate the halo mass range within which galaxies contain on average more (fewer) than $100$ stellar particles. 
Indicative-power law scalings for the mass loading factor are shown by the diagonal dashed black lines.
}
\label{mass_loading}
\end{center}
\end{figure*}

Fig.~\ref{mass_loading} presents the main results of this study, showing outflow rates for gas leaving the ISM (top panels)
and the halo (bottom panels) of central galaxies. Data are taken from the $100 \, \mathrm{Mpc}$ reference run, using trees with 200 snipshots.
Unless otherwise stated, all subsequent results in this paper are shown for this simulation using these trees.
Results are shown here as a function of halo mass; we refer readers interested in the dependence on
more readily observable quantities to Section~\ref{obsx}, where we show outflow rates as functions of stellar mass,
star formation rate, and circular velocity. We focus on central galaxies to simplify the interpretation of outflows
(which for satellites can also be caused by stripping by gravitational tides or gaseous ram pressure).

Following \cite{Neistein12}, the average measurements shown in Fig.~\ref{mass_loading} (and later figures) are taken by computing
the mean of the numerator over the mean of the denominator, including all central galaxies recorded within the quoted
redshift range. As demonstrated by \cite{Neistein12}, this approach yields the correct average mass exchange rate,
in the sense that taking the time integral over the averaged inflow and outflow rates predicts the correct stellar
masses of individual galaxies to within $0.1 \, \mathrm{dex}$ (because the mean of the time derivative of
the mass is equal to the time derivative of the mean of the mass). Taking the mean in this way also helps to
average out the discreteness noise that would affect outflow rate measurements of individual
galaxies if the numbers of outflowing particles and new stars formed between two simulation outputs is small.
We indicate with dashed lines the halo mass range where galaxies contain on average fewer than $100$
  stellar particles, which we take as an indicator of the range in which galaxies are poorly resolved.
  We show in Section~\ref{Convergence_sec} that this approximately corresponds to the mass scale above which our
  results are reasonably well converged with respect to numerical resolution.

%Note that at low halo masses a non-negligible fraction of galaxies do not form any star particles 
%over the redshift intervals shown, due to the finite resolution of the simulation. 
%We indicate mass bins where more than $20 \%$ of the galaxies have zero star formation with dashed lines, 
%as we expect this to indicate the range where the simulation results are definitely not converged 
%\cite[see][who show that the fraction of star forming versus passive galaxies is not converged for low-mass galaxies in \eagle]{Furlong15}. 

The left panels of Fig.~\ref{mass_loading} show average outflow rates normalised by the average star formation rates
computed over the same time interval (computed as the total mass of stars formed over the interval, ignoring
mass loss from stellar evolution). This quantity represents a time-averaged dimensionless mass loading factor, $\eta$
which can be considered as the efficiency with which outflows are launched from galaxies (top-left) and haloes (bottom-left).
Parametric fits to the mass loading factors are provided in Appendix~\ref{ap_parafit}.

Strong trends with halo mass are visible at both spatial scales, with a local minimum efficiency for outflows found at a halo mass around 
$M_{200} \sim 10^{12} \, \mathrm{M_\odot}$, approximately independent of redshift.
Below this characteristic halo mass, the galaxy-scale wind mass loading scales approximately as $M_{200}^{-0.5}$ (
the parametric best-fit value of the exponent is $-0.39 -0.06 \, z$), putting the % best-fit is -0.39 -0.06 z, so -0.4 at z=0, -0.5 at z=1, and -0.6 by z=3
\eagle simulations somewhere in between the often considered momentum-conserving ($\eta \propto V_{\mathrm{c}}^{-1} \propto M_{200}^{-1/3}$, 
where $V_{\mathrm{c}} \equiv \sqrt{G M_{200} / R_{\mathrm{vir}}}$ is the halo circular velocity) and energy-conserving scalings
($\eta \propto V_{\mathrm{c}}^{-2} \propto M_{200}^{-2/3}$). Note that these scalings only are only strictly
\emph{kinetic} energy and momentum conserving if the outflow velocity scales linearly with the circular
velocity of the system, which we show later is generally not the case for \eagle.
The corresponding mass loading scaling is typically steeper for the halo-scale outflows 
in the same mass range, with a best-fit exponent of $-1.19+0.18 \, z$, matching the
energy-conserving scaling ($\propto M_{200}^{-2/3}$) by $z \approx 3$.
Note that the scaling steepens noticeably for the galaxy-scale mass loading in the mass range where more than $20 \%$ of the galaxies are
not forming stars (indicated by dashed lines). This change in scaling towards very low mass may be therefore be related to 
resolution (and we typically exclude these mass bins from our analysis).

For $M_{200} > 10^{12} \, \mathrm{M_\odot}$, the mass loading factors start to rise again due to the effects of
AGN feedback (we show the explicit comparison with the no-AGN case in Section~\ref{agn_impact}).
The mass loading factor then declines slightly again for $M_{200} > 10^{13} \mathrm{M_\odot}$
for the galaxy-scale outflows, while the mass loading continues to rise monotonically with mass in high-mass
haloes for halo-scale outflows for $z<1$.
Put together, it is clear qualitatively that the scaling of the mass loading factors with halo mass is at least partly
responsible for the level of agreement between \eagle and the observed galaxy stellar mass function.
The scaling mimics the form of the empirically inferred relationship between $M_\star / M_{200}$ and $M_{200}$ \cite[e.g.][]{Moster18,Behroozi19},
in the sense that the maximum value of $M_\star / M_{200}$ is achieved at approximately the same halo mass where galactic outflows are least efficient
(per unit star formation).
% We stress that this behaviour is not simply inserted by hand into the subgrid modelling.
% Referee didn't like this sentence - can I dilute it a bit? - I think we can just low-key remove it - it is not needed

In the simplistic scenario where outflows alone set the scaling between stellar mass and halo mass,
the basic expectation is that $M_\star \propto \eta^{-1} \, M_{200}$, where $\eta$ is the mass loading factor \cite[][]{Mitchell16}.
Taking the example of the low-mass regime (where stellar feedback is typically assumed to dominate), empirical constraints
indicate the scaling between stellar mass and halo mass is approximately $M_\star \propto \, M_{200}^2$
\cite[e.g.][]{Behroozi19}, implying $\eta \propto M_{200}^{-1}$.
This is a stronger dependence compared to what we find in \eagle for galaxy-scale outflows, but is consistent
(particularly at lower redshifts) with the scaling we find for halo-scale outflows. This implies
first that at the spatial scale of galaxies, additional sources of mass scaling must be at play in order to match
the observed galaxy stellar mass function. The scaling of the halo-scale outflows could in principle be
a sufficient explanation (in that they reduce the available reservoir of baryons within the virial radius
that can accrete onto the ISM).
We defer a more quantitative analysis to a future study where we will present the corresponding picture
for gaseous inflows, which is required to fully understand the predicted relationship between stellar mass
and halo mass. 

The right panels of Figure~\ref{mass_loading} show outflow rates without normalizing by the star formation rates,
instead normalizing by halo mass to remove the zeroth order mass scaling to compress the dynamic range.
Starting with galaxy-scale outflows (top-right panel), it is interesting to note that the mass scale 
($M_{200} \sim 10^{12} \, \mathrm{M_\odot}$) where outflows are least
efficient in terms of the mass loading factor is where outflows are most efficient in terms of the
mass ejected per unit halo mass. This inversion serves to underline the aforementioned point that
the scaling between stellar mass and halo mass is stronger than that between galaxy-scale outflow rate and halo mass,
%(this can be seen in another way in the top-right panel of Fig.~\ref{outflow_mstar_sfr}), % - what did Joop mean here?
implying there must be other reasons for the stellar-halo mass scaling. %It is also evident in the
%top-right panel of Figure~\ref{mass_loading} that the mass scale where the ISM outflow rate per unit halo
%mass is maximimal evolves slightly with redshift, decreasing from $M_{200} \sim 10^{11.8} \, \mathrm{M_\odot}$ at $z=0$
%to $M_{200} \sim 10^{11.3} \mathrm{M_\odot}$ by $z=3$. Such an evolution is not as clearly seen in the ISM mass loading,
%implying that the evolution is related to an evolution in the efficiency of stellar assembly, rather than
%being tied to the efficiency of galactic outflows.
%%%%%%%%%% This part partially contradicts the parametric fit!
The picture changes markedly when considering instead the halo-scale outflow rates shown in the lower-right panel of Figure~\ref{mass_loading}.
The halo-scale outflow rates per unit halo mass are almost independent of halo mass for 
$M_{200} \sim 10^{10.5} - 10^{12.5} \, \mathrm{M_\odot}$, and for $z<1$ even up to $10^{14.5} \, \mathrm{M_\odot}$.
%A mild mass dependence
%starts to appear at lower redshifts, with the least mass of gas ejected out of haloes per unit halo mass
%for haloes of $M_{200} \sim 10^{11} \, \mathrm{M_\odot}$.

Differing degrees of redshift evolution at fixed halo mass can be seen in each panel of Figure~\ref{mass_loading}.
The galaxy-scale mass loading factor (top-left) decreases by about $0.5 \, \mathrm{dex}$ between $z=3$ and $z=0$
for haloes of mass, $M_{200} = 10^{11} \, \mathrm{M_\odot}$. We note that the respective positive and negative scalings of energy
injected by stellar feedback with gas density and metallicity \cite[Eqn~\ref{Eagle_eff}, see also figure 1 of][]{Crain15} could contribute to this
redshift evolution, as ISM densities/metallicities increase/decrease respectively with redshift at fixed 
mass.
Interestingly, the redshift dependence is reversed for the halo-scale mass loading factor (bottom-left
panel), with the efficiency of halo-scale outflows per unit star formation growing towards low redshift.
This presumably reflects an evolution of the properties of circum-galactic gas out to the virial radius.
Another possibility is that halo-scale outflows are being driven by energy injected in the past, when star formation rates were higher. % Could be related to a change in potential well as f(z) at fixed halo mass.

Considering instead the outflow rates normalized by halo mass (right panels) instead of by star formation rate, 
a trend of outflow rates increasing with increasing redshift is apparent for both galaxy and halo-scale
outflows. This primarily reflects the evolution of galaxy star formation rates at fixed halo mass, which 
in turn is related to the slowing of structure formation towards low redshift that occurs in the $\Lambda$CDM 
cosmological model. Indeed, if the outflow rates shown in the right panels are multiplied by the age of the Universe
for each redshift bin (in effect removing the redshift scaling of dark matter halo accretion rate), most of the 
redshift evolution disappears for the galaxy-scale outflows, and almost all of the redshift evolution disappears 
for the halo-scale outflows.

\begin{figure}
\includegraphics[width=20pc]{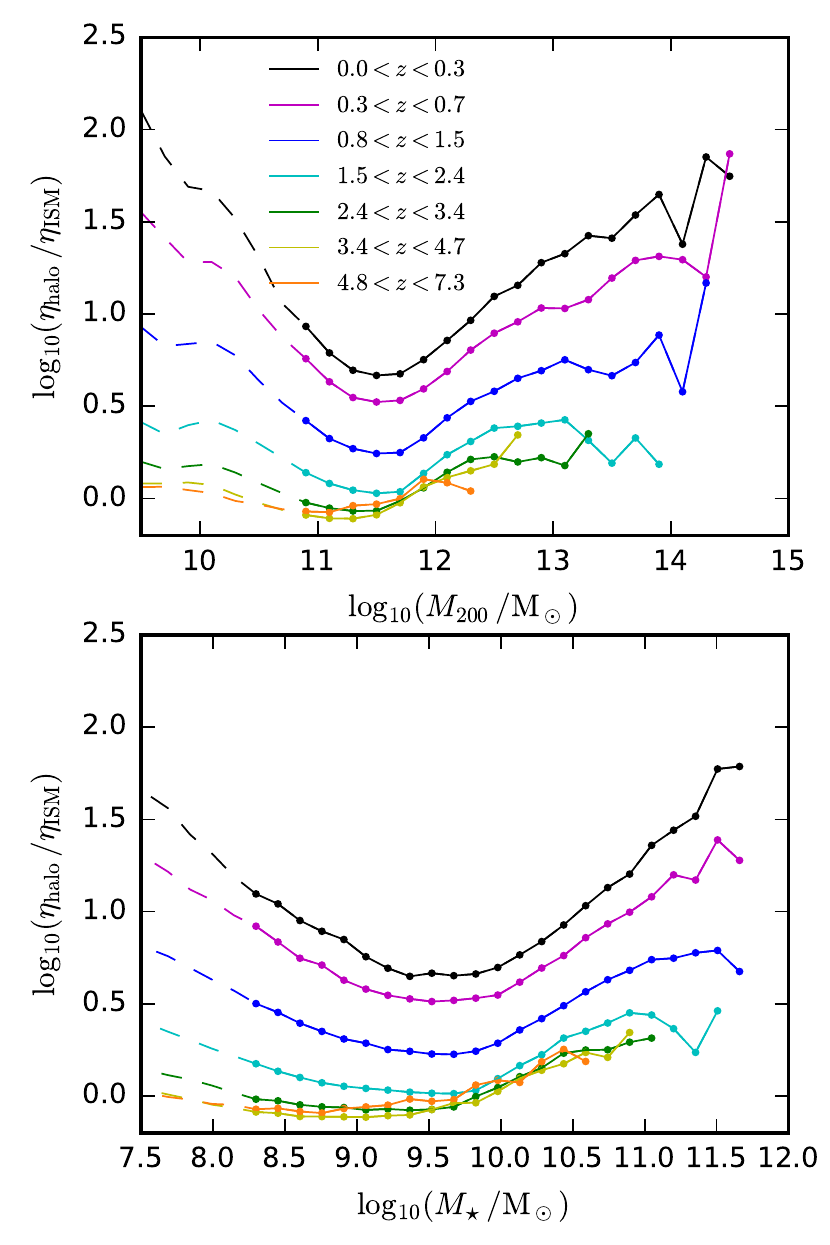}
\caption{The ratio of halo-scale mass loading factor to galaxy-scale mass loading factor, plotted as a function of halo mass (top panel), and stellar mass within a $30 \, \mathrm{pkpc}$ spherical aperture (bottom panel).
Solid (dashed) lines indicate the halo mass range within which galaxies contain on average more (fewer) than $100$ stellar particles.
In general, substantially more outflowing mass is being removed from the halo than is being removed from the ISM.
%Data are taken from the $100 \, \mathrm{Mpc}$ reference run, using trees with 200 snipshots.
%\textbf{As a placeholder, data are taken from the $25 \, \mathrm{Mpc}$ reference run, using trees with 200 snipshots.}
}
\label{mass_loading_ratio}
\end{figure}

\begin{figure*}
\begin{center}
\includegraphics[width=40pc]{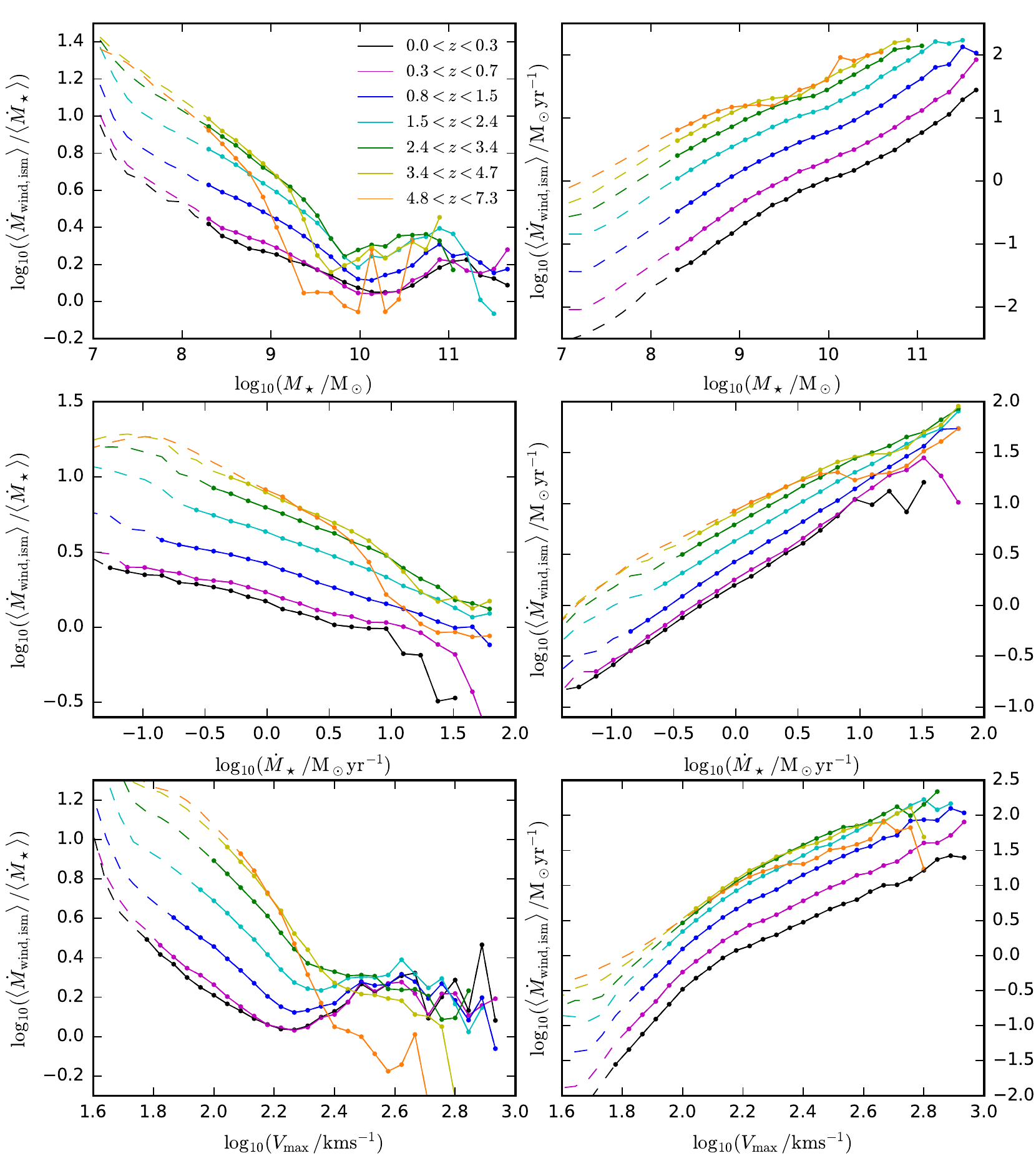}
\caption{Galaxy-scale outflow rates as functions of stellar mass, $M_\star$ (top), star formation rate, $\dot{M}_\star$ (middle), and halo maximum circular velocity, $V_{\mathrm{max}}$ (bottom).
Left panels show the average mass loading factor plotted as a function of different variables, and right panels show the average outflow rate.
Solid (dashed) lines indicate bins within which galaxies contain on average more (fewer) than $100$ stellar particles.
}
\label{outflow_mstar_sfr}
\end{center}
\end{figure*}

\subsection{Comparing outflow rates at galaxy and halo scales}

An important feature of the rates shown in Figure~\ref{mass_loading} is that in general, substantially more mass
is flowing out of the halo virial radius compared to that leaving the ISM. We show this
explicitly in Fig.~\ref{mass_loading_ratio}. At high redshift ($z>3$), the halo and galaxy-scale
outflow rates are roughly equal for halo masses $M_{200} < 10^{12} \, \mathrm{M_\odot}$ (or for $M_\star < 10^{10} \, \mathrm{M_\odot}$).
For $z<2$, the halo-scale outflow rates evolve to become increasingly elevated over the galaxy-scale rates
at lower redshift. The mass dependence becomes stronger at lower redshifts,
with halo-scale outflows becoming increasingly elevated over galaxy-scale outflows in both low-mass and high-mass
haloes, transitioning around a minimum elevation at $M_{200} \sim 10^{11.5} \, \mathrm{M_\odot}$.

All together, the enhanced outflow rates at the halo virial
radius will play an important role in the \eagle simulations, by
effectively reducing the reservoir of baryons within the
virial radius that can condense down onto the ISM,
but without invoking outflow rates at the galaxy scale that are far too
high relative to observational constraints (see Section~\ref{obs_comp_sec}).
We explore the origins of the enhancement in the following
parts of this section, culminating in the discussion presented
in Section~\ref{Entrainment_discussion}.
The question of whether of the mass loading factors at
the two scales are qualitatively and quantitatively robust
with respect to changing numerical resolution is discussed in Section~\ref{Convergence_sec}.

\subsection{Outflow rates as functions of $M_\star$, $\dot{M}_\star$, and $V_{\mathrm{max}}$}
\label{obsx}

Fig.~\ref{outflow_mstar_sfr} shows galaxy-scale outflow rates as functions of stellar mass, $M_\star$, star formation rate,
$\dot{M}_\star$, and halo maximum circular velocity, $V_{\mathrm{max}}$, quantities that are more readily 
observable than halo mass.
For outflow rates plotted as a function of $\dot{M}_\star$, galaxies are binned according to the mass of stars formed within the last $100 \, \mathrm{Myr}$,
comparable with the characteristic time-scale of SFR measurements derived from UV luminosities, but to be self-consistent the star formation 
rate folded into the mass loading factor is always taken from the mass of stars that formed within the same time interval used to measure the outflow rate.
The stellar masses and star formation rates plotted along the x-axis are both measured using only star particles within a $30 \, \mathrm{pkpc}$ spherical aperture.
Parametric fits for the mass loading factor as a function of $M_\star$ and $V_{\mathrm{max}}$ are given in Appendix~\ref{ap_parafit}.

While trends are similar to those seen in Fig.~\ref{mass_loading}, several notable features do stand
out in Fig.~\ref{outflow_mstar_sfr}. While the scaling of galaxy-scale outflow rates plotted as a function of halo mass 
(upper-right in Fig.~\ref{mass_loading}) or maximum circular velocity (bottom-right in 
Fig.~\ref{outflow_mstar_sfr}) show a characteristic change in slope around $M_{200} \sim 10^{12} \, \mathrm{M_\odot}$
or $V_{\mathrm{max}} \sim 125 \, \mathrm{km s^{-1}}$, such a change is much less evident in the scaling
of outflow rate with stellar mass (top-right Fig.~\ref{outflow_mstar_sfr}). 
This difference reflects in combination the mass scaling of the mass loading factor, the dependence of star formation rate
per unit stellar mass on stellar mass \cite[see figure 5 in][]{Furlong15}, and the underlying scaling
of galaxy stellar mass on halo mass \cite[see figure 8 in][]{Schaye15}.

Another feature visible in Fig.~\ref{outflow_mstar_sfr} is that the negative scaling of the mass loading factor
with star formation rate (middle-left) does not flatten or turn over for high star formation rates, unlike for all of the other
variables considered. This reflects the strong decrease of galaxy star formation rates per unit stellar mass
in massive galaxies (where AGN power most of the outflow and so change the mass scaling of the mass loading factor, see section~\ref{agn_impact}), such that massive
galaxies do not dominate the highest star formation rate bins.

\subsection{Outflow velocities}

\begin{figure}
\includegraphics[width=20pc]{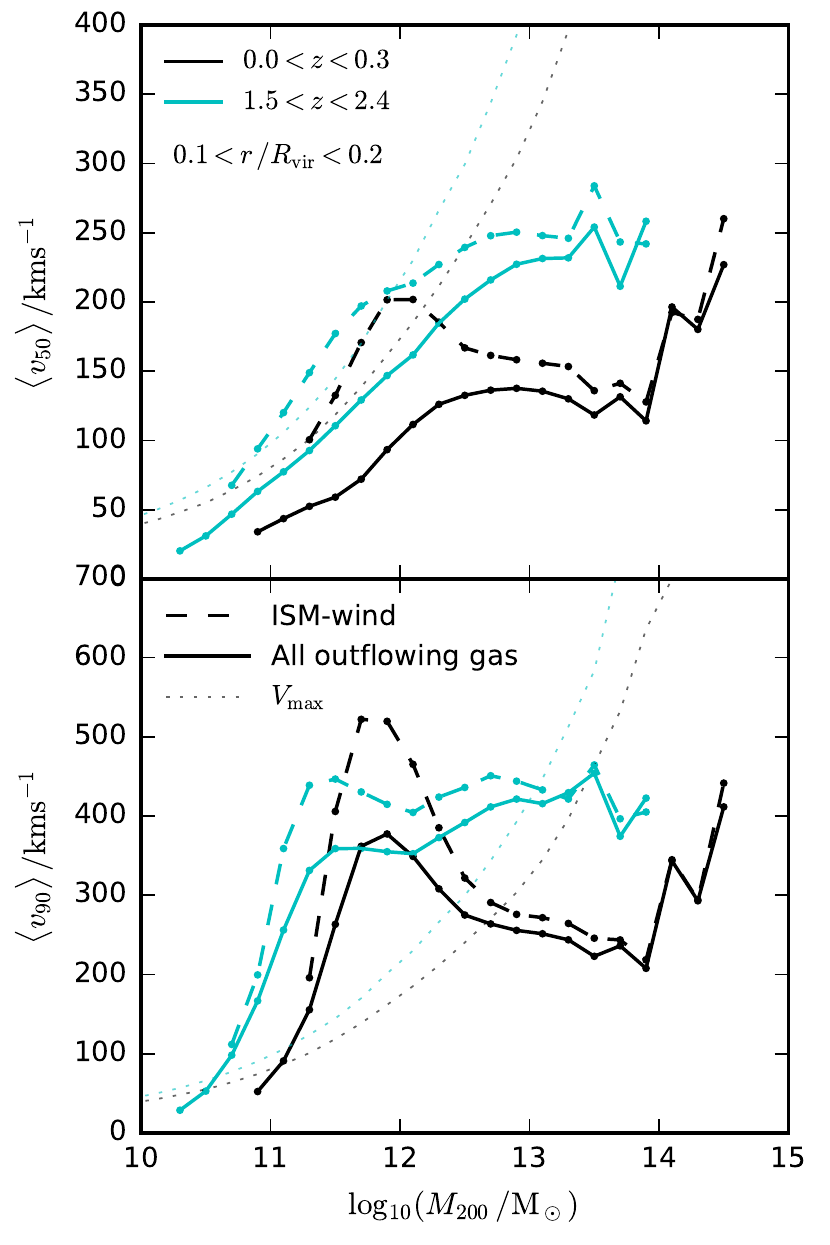}
\caption{
The median (mass) flux-weighted radial velocity of all outflowing gas (solid lines), and outflowing gas that has been ejected from the ISM (dashed lines), plotted as a function of halo mass.
Flux-weighted velocities for each galaxy are computed as either the median weighted velocity ($v_{\mathrm{50}}$, top panel), or the $90^{\mathrm{th}}$ percentile ($v_{\mathrm{90}}$, bottom panel).
Velocities are measured in spherical shells at radius $0.1 < r \, / R_{\mathrm{vir}} < 0.2$. 
The median relationship between maximum halo circular velocity, $V_{\mathrm{max}}$, and halo mass is also shown (dotted lines).
Outflow velocities are only shown for halo mass bins where more than $80 \%$ of the galaxies have non-zero flux within the shell.
Median outflow velocities are in general comparable to $V_{\mathrm{max}}$ for $M_{\mathrm{200}} < 10^{12} \, \mathrm{M_\odot}$, 
but saturate (or even decline in some cases) for high-mass haloes.
}
\label{velocity_ism_wind}
\end{figure}

\begin{figure*}
\begin{center}
\includegraphics[width=40pc]{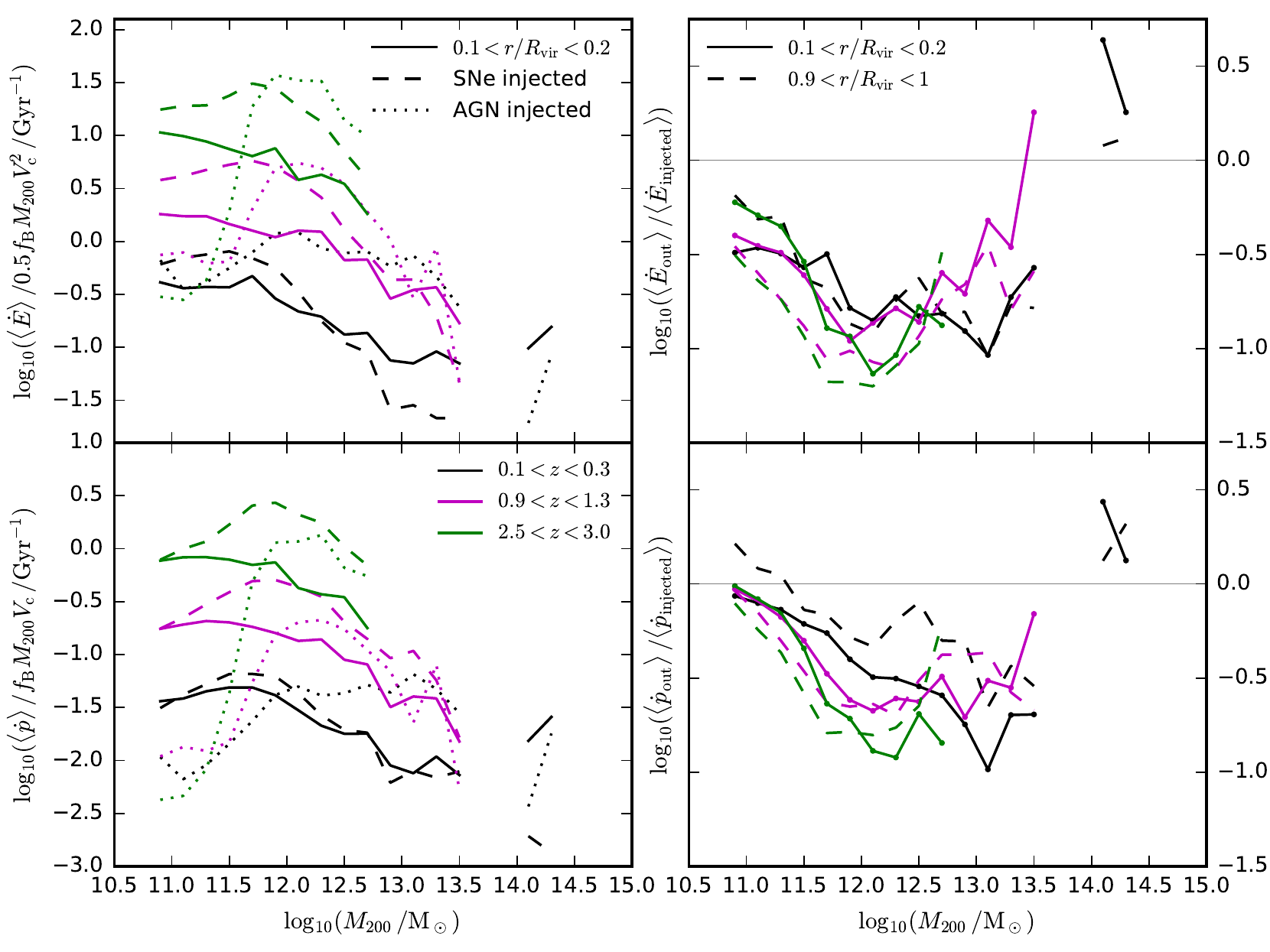}
\caption{Energy (thermal plus kinetic, top) and radial momentum (bottom) fluxes of outflowing gas, plotted as a function of halo mass.
\textbf{\textit{Left panels:}} solid lines show fluxes of outflowing gas ($v_{\mathrm{rad}} > 0 \, \mathrm{km s^{-1}}$) within a spherical shell ($0.1<r/R_{\mathrm{vir}}<0.2$).
These can be compared to the input thermal energy injection rate or (pseudo-)input momentum from stellar (dashed lines) and AGN feedback (dotted lines).
Because feedback in \eagle is purely thermal, the (pseudo-)input momentum rate is defined relative to the thermal energy injection rate as
%$\dot{p}_{\mathrm{injected}} = \sqrt{2 \dot{E}_{\mathrm{inject}} \dot{M}_{\mathrm{heated}}}$, % Error in the first version
$\dot{p} = \frac{1}{\Delta t} \, \sqrt{2 \Delta{E}_{\mathrm{inject}} \Delta{M}_{\mathrm{heated}} }$
  where $\Delta E_{\mathrm{inject}}$ is the energy that is directly injected into $\Delta M_{\mathrm{heated}}$ of mass over a time interval $\Delta t$ 
 (see main text).
Fluxes and injection rates are normalised by the characteristic energy/momenta of the associated haloes.
\textbf{\textit{Right panels:}} fluxes of outflowing gas divided by the corresponding energy/(pseudo-)momentum injection rates, defining effective energy or momentum loading factors.
Loading factors are shown for outflowing gas in shells at $0.1<r/R_{\mathrm{vir}}<0.2$ (solid lines), and at $0.9<r/R_{\mathrm{vir}}<1.0$ (dashed lines).
In all panels, data are only shown for mass bins within which galaxies contain on average more than $100$ stellar particles.
Gas within the ISM is excluded from the flux measurements.
Data are taken from the $50 \, \mathrm{Mpc}$ reference run.
Roughly $20 \%$ of the energy being injected by feedback is retained in outflows in \eagle for $M_{200} \sim 10^{12} \, \mathrm{M_\odot}$, with this fraction increasing for both higher and lower halo masses.}
\label{energy_momentum}
\end{center}
\end{figure*}

While the main focus of this study is on outflow rates, it is also interesting to explore
the decomposition of these gas flows as a function of velocity, or gas phase. We defer a detailed
analysis to future work, but we do show here the average flux-weighted velocity of 
outflowing gas in Fig.~\ref{velocity_ism_wind}. The median velocities (top panel) exhibit
roughly logarithmic scaling with halo mass. Outflowing gas that was ejected from the ISM
moves at higher velocities relative to all outflowing gas at a given radius, and exhibits
a peak velocity at a characteristic halo mass of $10^{12} \, \mathrm{M_\odot}$ at $z=0$.
This effect is more pronounced for the $90^{\mathrm{th}}$ percentile of the flux-weighted
outflow velocity (bottom panel). 
Except for the scaling of median velocity with halo
mass in low-mass haloes ($M_{200} < 10^{12} \, \mathrm{M_\odot}$), the scaling of outflow velocity is 
qualitatively different to the scaling of maximum halo circular velocity with halo mass (shown by the dotted lines).
The spread in velocities at a given mass/redshift is large (as can be appreciated by
comparing the two percentiles). Outflow velocities at a given halo mass are higher
at higher redshifts, with the exception of $v_{\mathrm{90}}$ around the peak at
$M_{200} \sim 10^{12} \, \mathrm{M_\odot}$.

% Another relevant point here is that most of the ISM wind is able to escape the halo, despite having typical velocities that are too low for massive haloes
% This indicates that the trajectories are definitely not balistic
% Annoyingly, I can't really make this point (it may even be wrong) because of the bug with the fraction of ISM wind leaving the halo plot.

\subsection{Energy and momentum fluxes}
\label{energy_momentum_sec}

\begin{figure*}
\begin{center}
\includegraphics[width=40pc]{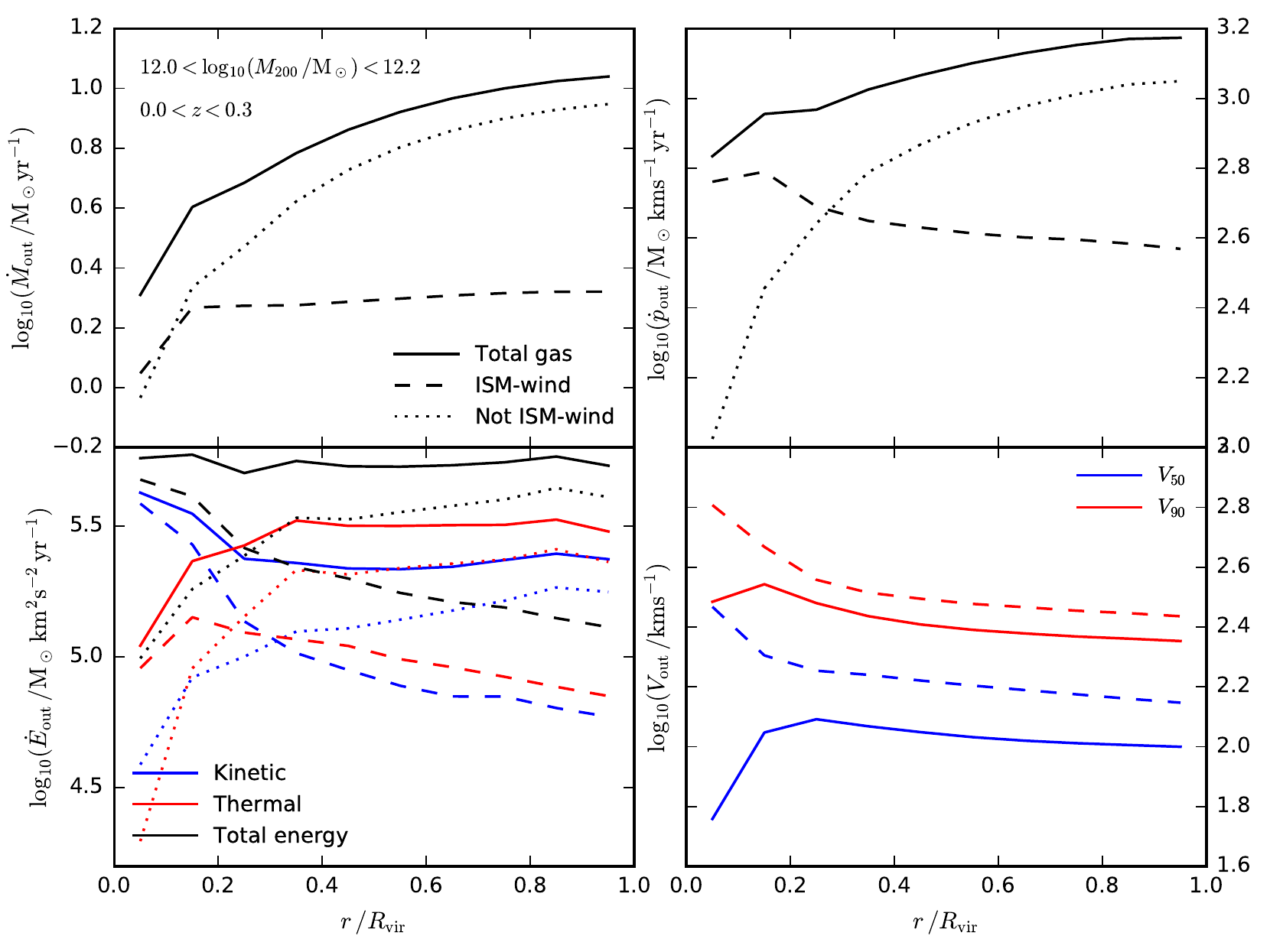}
\caption{Mean mass (top-left), radial momentum (top-right), and energy (bottom-left) fluxes plotted as
a function of radius for haloes with mass $12 < \log_{10}(M_{200} \, / \mathrm{M_\odot}) < 12.2$ for redshift $0 < z < 0.3$.
Solid lines show these quantities for all outflowing gas ($v_{\mathrm{rad}} > 0$), dashed lines show outflowing gas
identified as part of the wind that left the ISM, and dotted lines show the remaining outflowing gas
(note that the latter selection is not computed for the outflow velocity, bottom-right panel).
Gas within the ISM is excluded from the flux measurements.
The bottom-right panel shows the mass flux-weighted $50^{\mathrm{th}}$ and $90^{\mathrm{th}}$ percentiles of the
distributions of radial velocity (for the same selections of gas).
Outflowing mass and momentum fluxes rise as winds propagate outwards for this halo mass and redshift range,
while the energy flux remains approximately constant, with energy
seemingly being transferred from the material ejected from the ISM to the ambient halo gas.
}
\label{outflow_shells}
\end{center}
\end{figure*}

While the mass loading factor of galactic winds is one measure of their efficiency, it is also interesting to assess
the wind efficiency in terms of energy and radial momentum. Fig.~\ref{energy_momentum} shows measurements of
the fluxes of energy (kinetic plus thermal) and momentum, contrasted with the rate of thermal energy injection by feedback processes ($\dot{E}_{\mathrm{inject}}$).
While zero momentum is injected by hand in the simulation, we can define an effective momentum injection rate
as $\frac{\Delta p}{\Delta t} = \frac{1}{\Delta t} \, \sqrt{2 \Delta{E}_{\mathrm{inject}} \Delta{M}_{\mathrm{heated}} }$,
where $\Delta E_{\mathrm{inject}}$ of energy is directly injected into $\Delta M_{\mathrm{heated}}$ of mass
over a time interval $\Delta t$. This represents the momentum that the wind
would achieve if all of the injected thermal energy were converted to kinetic form, and should be
regarded as a rule of thumb rather than as the true momentum that winds are expected to attain. %Note that
%this is not a converged quantity; in reality supernova remnants carry much less mass per unit energy than the mass that is
%directly heated in the simulation, and so the true input momentum would accordingly be lower at fixed energy.}

The top-left panel of Fig.~\ref{energy_momentum} shows the energy flux of outflowing gas close to the galaxy
(solid lines), normalised by the kinetic energy that would be required to move the entire
baryonic content of the halo at the halo circular velocity, $V_{\mathrm{c}}$, assuming the baryon to dark matter content
of the halo matches the universal fraction, $f_{\mathrm{B}}$. At high redshift, more than sufficient energy
is being injected to achieve this within a $\mathrm{Gyr}$, but this is no longer the case at low redshift
once the rates of star formation and SMBH accretion have slowed at fixed halo mass. The upper-right panel shows
the ratio of the energy flux to the feedback energy injection rate, both close to the galaxy (solid lines)
and at the virial radius of the halo (dashed lines). While these measurements are noisier than for the
mass loading factor\footnote{Energy fluxes are noisier because we have to perform measurements in discrete shells, and because a relatively small number
of particles can carry a high fraction of the outflowing energy.}, the trend of energy loading with mass qualitatively matches that
of the mass loading, with a minimum value at $M_{200} \sim 10^{12} \, \mathrm{M_\odot}$.
Outflows contain about $30 \%$ of the injected energy at $M_{200} = 10^{11} \, \mathrm{M_\odot}$,
which drops to about $10 \%$ at $M_{200} = 10^{12} \, \mathrm{M_\odot}$. 

At low ($M_{200} \approx 10^{11} \, \mathrm{M_\odot}$) and high ($M_{200} > 10^{13.5} \, \mathrm{M_\odot}$) halo masses,
the outflows can carry more energy than is being injected. This serves
first to underline that the energy loading factors plotted are upper limits to the efficiency with
which the injected energy from feedback is able to power galactic winds. Other sources of
energy in outflowing gas include the ultraviolet background (UVB, which could plausibly be responsible 
for the greater than unity energy loading measured for outflows at the virial radius in low-mass haloes),
and gravitational heating (which could plausibly have a larger relative effect in massive
haloes, where pressurized hot coronae have developed). Another factor is
that the energy/momentum fluxes at the halo virial radius are associated with
feedback events that occurred earlier in the history of each galaxy, at which
time the star formation and SMBH accretion rates may have been significantly different.
We return to this point in Section~\ref{Entrainment_discussion}.

For intermediate-mass haloes, the energy in outflows close to the galaxy is typically higher
than for outflows close to the virial radius, likely indicating dissipation over the intervening
scales. This is less apparent when comparing the momentum flux at the two scales, and by $z=0$
the momentum flux is higher at the virial radius than near the galaxy over the entire halo mass range probed (other
than the handful of haloes in the highest mass bins). This indicates some level of entrainment 
of mass at fixed energy, which is consistent with the enhanced mass loading at the virial
radius seen in Fig.~\ref{mass_loading_ratio}.

\subsection{Outflows as a function of radius}
\label{relationship_sec}

Entrainment of outflowing mass is shown more directly in Fig.~\ref{outflow_shells}, which
shows the mass, momentum and energy fluxes as a function of radius for haloes of mass $12 < \log_{10}(M_{200} \, / \mathrm{M_\odot}) < 12.2$
for redshifts $0 < z < 0.3$. In this instance, we separate the contribution from gas that has been
removed from the ISM (dashed lines), versus gas that has has never been in the ISM (dotted lines).
Mass flux (top-left panel) is conserved as a function of radius 
for the former ISM material, but by $0.2 \, R_{\mathrm{vir}}$ there is a similar
mass flux of material that was never in the ISM, and the contribution of this component
rises until it dominates the mass flux at the virial radius. A similar picture is
seen for the momentum flux (top-right panel). 

The total energy flux (solid black line in
the bottom-left panel) is approximately constant with radius, with energy seemingly
being exchanged from the former ISM component (dashed black line) to gas entrained
from the circum-galactic medium (dotted black line) as outflows propagate outwards.
Despite the feedback scheme employed in \eagle being thermal, the majority of the outflowing energy
flux is in kinetic form close the galaxy, but the majority of the energy flux is in thermal
form at larger radii. Correspondingly, the mass flux-weighted velocities (bottom-right panel)
decline as a function of radius.

%how can the ISM wind mass flux be constant with radius but the momentum flux drop (slightly) with radius? 
%This can happen if we lose mass from the ISM wind, but gain velocity, keeping the mass flux constant.
% Because the momentum flux scales with velocity squared.
% hmm, but the flux-weighted velocities drop with radius. - although that is just a moment of the distribution I guess

Overall, the trends are consistent with a picture whereby gas is entrained on circum-galactic
scales, explaining much of the difference between the halo and galaxy-scale outflow rates
shown in Fig.~\ref{mass_loading_ratio}. A similar picture is seen at lower halo masses 
at low redshift (not shown), although in that instance the total energy flux actually 
rises with radius, indicating another source of energy is involved (possibly the UVB). The picture
is again similar at higher halo masses, but in this case the entrainment phenomenon
ceases once the outflow reaches half the halo virial radius, thermal energy is
more dominant over kinetic energy, and the fractional contribution to the energy flux from outflowing material that
has never been in the ISM is higher at the center. At higher redshifts, the trends are
similar but there is systematically less evidence for entrainment, as the mass flux
increases much less strongly with radius (as seen also in Fig.~\ref{mass_loading_ratio}).

\subsection{Directionality}
\label{directionality}

\begin{figure}
\includegraphics[width=20pc]{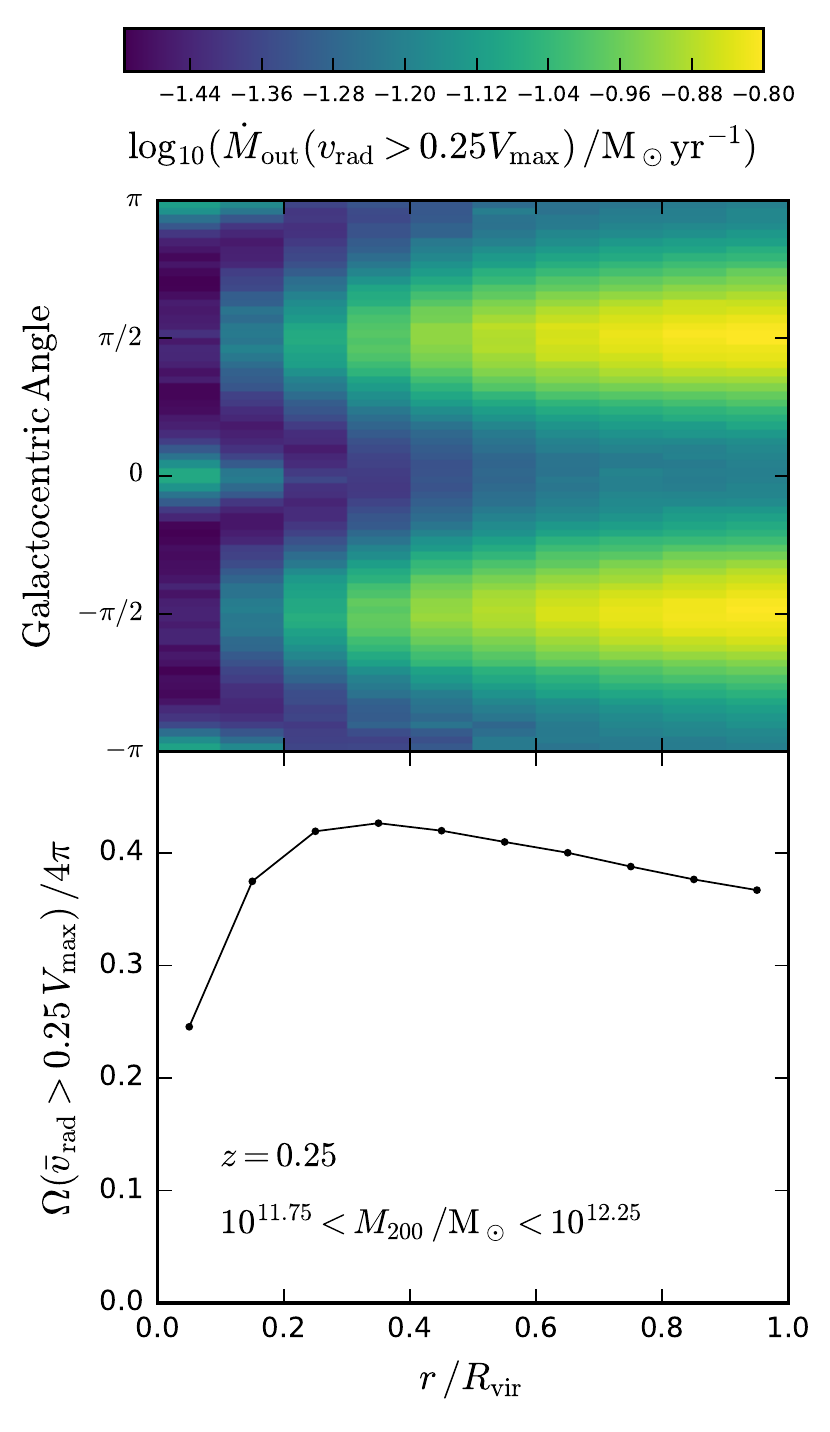}
\caption{Directionality (top) and spherical solid-angle covering fraction (bottom) of
  outflowing gas, plotted as a function of radius,
  averaging over galaxies with $10^{11.75} < M_{\odot} < 10^{12.25}$ at $z=0.25$.
  \textbf{\textit{Top:}} directionality is defined in terms of the galactocentric angle (see main text), with
  values of $\pm \pi /2$ indicating that gas is flowing orthogonally to the disk (minor axes).
  The colour scale indicates the radially outflowing mass flux, as labelled, selecting only
  gas particles with radial velocities $v_{\mathrm{rad}} > 0.25 V_{\mathrm{max}}$.
  For $r > 0.1 \, R_{\mathrm{vir}}$, \eagle produces a clear bimodal outflow distribution that aligns with the galaxy minor axis.
  \textbf{\textit{Bottom:}} the solid angle fraction of the sphere that
  is covered by solid angle bins within which the \emph{net} flux-weighted radial velocity
  satisfies $\langle v_{\mathrm{rad}} \rangle > 0.25 V_{\mathrm{max}}$.
  For $r > 0.2 \, R_{\mathrm{vir}}$ about $40 \%$ of the virial sphere is occupied by outflowing gas.
  }
\label{direct}
\end{figure}

  The top panel of Fig.~\ref{direct} shows an example of the angular dependence of galactic outflows
  in \eagle. Following the approach used in \cite{Nelson19}, we show mass flux as a function of
  radius (x-axis) and ``galactocentric'' angle, which we take as the angle between each gas particle
  and the major axis of the galaxy, as viewed in an edge-on projection
  (defining the mid-plane using the angular momentum vector of the ISM). For a disk-galaxy,
  values of $0$ and $\pm \pi$ therefore indicate outflows that are
  propagating within the plane of the disk, and values of $\pm \pi /2$ indicate outflows that
  propagate orthogonally to the disk along the minor axes.
  The increase of mass flux with radius shows once again the previously-discussed entrainment
  effect. The distribution with angle shows that \eagle produces a bimodal outflow pattern,
  aligned with the minor axes, which reflects the relative ease with which outflows can escape
  the ISM (and propagate through the CGM) in the directions orthogonal to the disk.
  Note that at $r<0.2 \, R_{\mathrm{vir}}$ there is also some outflowing flux aligned with the disk,
  which we interpret as a combination of ISM material (which was not subtracted here) and gas
  that is settling around the disk after infall.

  The bottom panel of Fig.~\ref{direct} shows as a function or radius the fraction of the virial sphere
  that is occupied by gas that is on average outflowing with
  $v_{\mathrm{rad}} > 0.25 V_{\mathrm{max}}$. The fraction rises from $\approx 20 \%$ at the halo
  center up to a peak value close to $40 \%$ at $r = 0.3 R_{\mathrm{vir}}$, and stays nearly
  constant out to larger radii. The enhancement of the mass flux with radius is therefore
  not associated with an increase in the solid angle of the outflow for $r > 0.3 R_{\mathrm{vir}}$.

\subsection{Energy-driven winds and travel-time effects}
\label{Entrainment_discussion}

Having presented information on the mass, momentum and energy fluxes, velocities, and
directionality of galactic outflows, we can now put this together to discuss the
origin of the enhancement in mass flux with radius (out to the virial radius) seen in \eagle.
We stress at the outset that this is a question that is complicated to address in cosmological
simulations because of evolution effects: galaxies and  haloes can grow significantly both in mass
and size over timescales that are comparable to the timescales for circum-galactic
gas flows, the velocity field of circum-galactic gas will also reflect cosmological infall,
and the energy and momentum content of circum-galactic gas at different scales will reflect the
cumulative injection of feedback energy over a range of timescales. We can nonetheless examine some
simplified arguments, which we present here.

An obvious mechanism to increase mass flux with radius comes from an ``energy-driven'' wind scenario,
in which the outflows are over-pressurized relative to the ambient ISM or CGM, which
generates radial momentum as the hot interior of the outflow does $P \mathrm{d}V$ work
on the surrounding gas. This is the physical mechanism responsible for increasing
the radial momentum of an outflow during the Sedov-Taylor (adiabatic) and pressure-driven snowplow
phases of supernova explosions. It has also been discussed within
the context of larger-scale AGN-drive winds \cite[e.g.][]{King11,FaucherGiguere12}, which
have been demonstrated to be capable of driving an increase of mass flux with radius on
circum-galactic scales in full cosmological simulations \cite[][]{Costa14}.

\begin{figure}
\includegraphics[width=20pc]{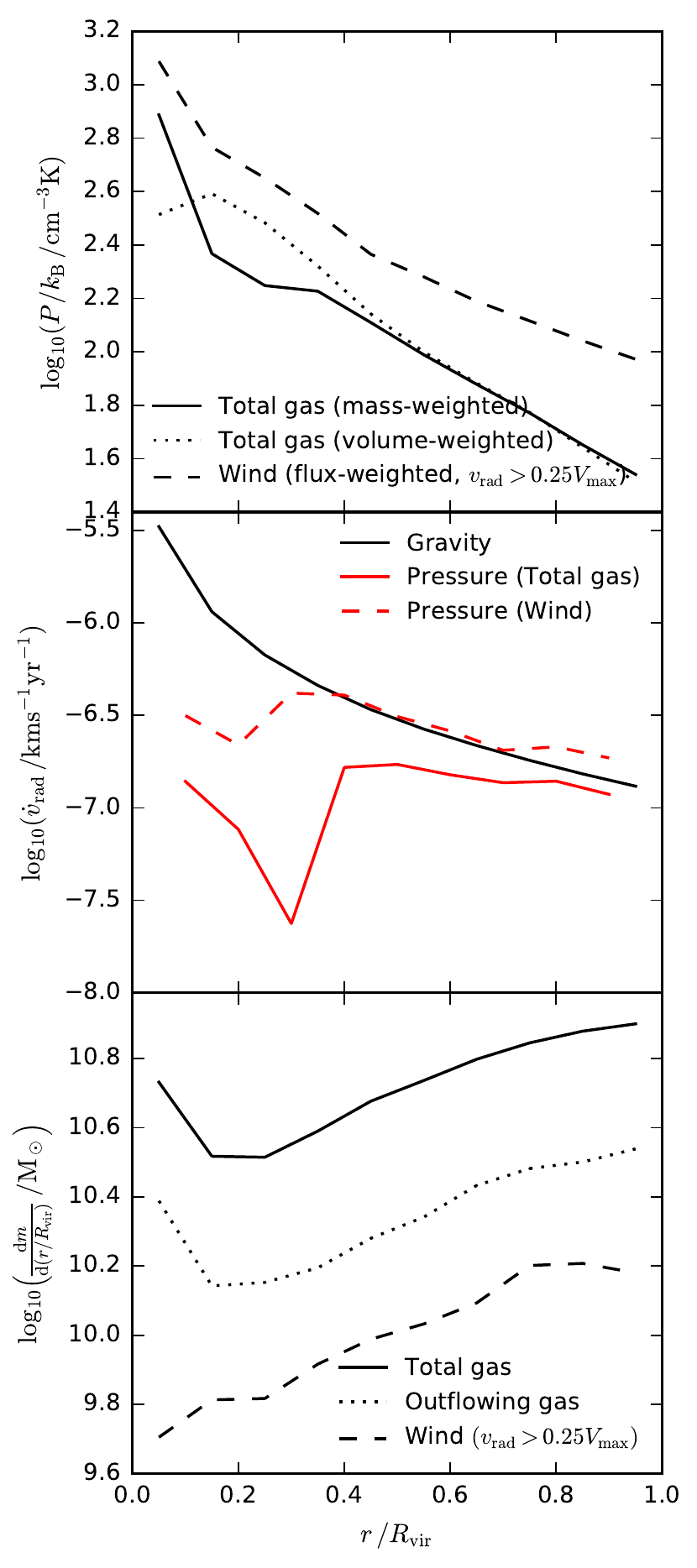}
\vspace*{-22pt}
\caption{
  Radial profiles of thermal pressure (top), radial acceleration (middle), and mass (bottom),
  averaged over galaxies with $10^{12} < M_{\odot} < 10^{12.2}$ at $z=0.25$. 
  \textbf{\textit{Top:}} median thermal pressure profiles for all gas, both
  mass-weighted (solid line) and volume-weighted (dotted line), as well as
  the flux-weighted median for outflowing gas with $v_{\mathrm{rad}} > 0.25 V_{\mathrm{max}}$ (dashed line),
  which we take as a measurement of the characteristic thermal pressure within winds. Winds are on average over-pressurized
  relative to the typical CGM at a given radius, which will drive an increase in momentum
  as outflows propagate outwards.
  \textbf{\textit{Middle:}} median radial acceleration due to the negative thermal pressure
  gradient (red lines) for all gas (mass-weighted, solid line) and for outflowing gas with
  $v_{\mathrm{rad}} > 0.25 V_{\mathrm{max}}$ (flux-weighted, dashed line). Also shown is the (inwards) radial
  acceleration associated with the gradient of the gravitational potential, assuming
  spherical symmetry. Gas in winds is on-average pressure supported against gravitational acceleration
  for $r > 0.4 \, R_{\mathrm{vir}}$, but on average gas within the CGM is under-supported at all radii.
  \textbf{\textit{Bottom:}} radial mass profiles for total gas (solid), all outflowing gas (dotted),
  and outflowing gas with $v_{\mathrm{rad}} > 0.25 V_{\mathrm{max}}$ (dashed). Most of the mass
  in the CGM is in the outer regions of haloes, helping to explain why wind entrainment acts on large
  spatial scales.
  }
\label{pressure_fig}
\end{figure}

  The top panel of Fig.~\ref{pressure_fig} shows the radial profile of
the median thermal pressure, averaging over galaxies with
$10^{12} < M_{\odot} < 10^{12.2}$ at $z=0.25$. We show
the estimates of the average thermal pressure of all gas 
at a given radius (solid/dotted lines, corresponding to mass/volume weighting)
as well as the flux-weighted average thermal pressure of outflowing gas with
$v_{\mathrm{rad}} > 0.25 V_{\mathrm{max}}$ (dashed line), which we take as
a measure of the characteristic thermal pressure within feedback-driven winds.
To compute the weighted average of $P_{\mathrm{gas}}$, we weigh by 
$m_{\mathrm{gas}} \, P_{\mathrm{gas}}$
for mass-weighted, $(m_{\mathrm{gas}} / \rho_{\mathrm{gas}}) \, P_{\mathrm{gas}}$
for volume-weighted, and $(m_{\mathrm{gas}} v_{\mathrm{rad}}) \, P_{\mathrm{gas}}$
for flux-weighted, where $m_{\mathrm{gas}}$ is the gas particle mass, $\rho_{\mathrm{gas}}$ is the SPH 
density, $v_{\mathrm{rad}}$ is the radial velocity, and $P_{\mathrm{gas}}$ is the SPH pressure.
We average over spherical shells of width $0.1 \, R_{\mathrm{vir}}$,
including particles whose centers are inside the shell.

We find that the outflowing gas is over-pressurized relative to the
total CGM by $\approx 0.3 \mathrm{dex}$ at all radii, which will drive
an increase of momentum with time and distance for discrete outflow
events as they propagate through the ambient CGM.
  Referring back to Fig.~\ref{outflow_shells}, which shows
  that the energy flux is roughly constant with radius for this
  mass/redshift range, it therefore appears that winds are
  driven across the CGM in an energy-driven configuration.
  Fig.~\ref{outflow_shells} also shows that the average wind velocity
  is nearly flat with radius (more precisely it is slightly
  declining), which implies that the increase in mass flux
  must be associated with entrainment of ambient gas, not
  with an increase in the characteristic velocity of the
  outflow with radius. The bottom panel of Fig.~\ref{direct}
  shows that this entrainment is not associated with
  an increase in the solid angle occupied by winds as
  a function of radius. Putting the information from
  these measurements together implies then that
  the mass per unit radius in outflows must increase
  with radius, which is shown explicitly to be the
  case in the bottom panel of Fig.~\ref{pressure_fig}
  \cite[see also][for a focussed analysis of density profiles
  in \eagle]{Schaller15b}.
  With a spherically averaged density profile that
  is shallower than isothermal (for which $\rho(r) \propto r^{-2}$),
  most of the mass in the CGM is in the outer regions
  of the halo in \eagle, which helps to explain why
  the entrainment effect is seen at larger scales.

  The middle panel of Fig.~\ref{pressure_fig} shows an estimate of the typical
  radial acceleration imposed by the radial thermal pressure gradient for gas in
  winds (dashed red line), and for all gas (solid red line). This is contrasted
  against the (opposite-sign) gravitational radial acceleration (black line).
  The CGM is on average under-supported against gravitational infall, and will
  therefore tend towards a net inflow in the absence of additional sources
  of inflow/outflow from the ISM and from beyond the virial radius
  \cite[see][for a generalised discussion of hydrostatic balance in the \eagle simulations]{Oppenheimer18}.
  Gas within winds is pressure supported against the gravitational acceleration
  for $r > 0.4 \, R_{\mathrm{vir}}$, explaining why the radial velocity of outflows
  (Fig.~\ref{outflow_shells}) is nearly flat in this radius range.

\begin{figure}
\includegraphics[width=20pc]{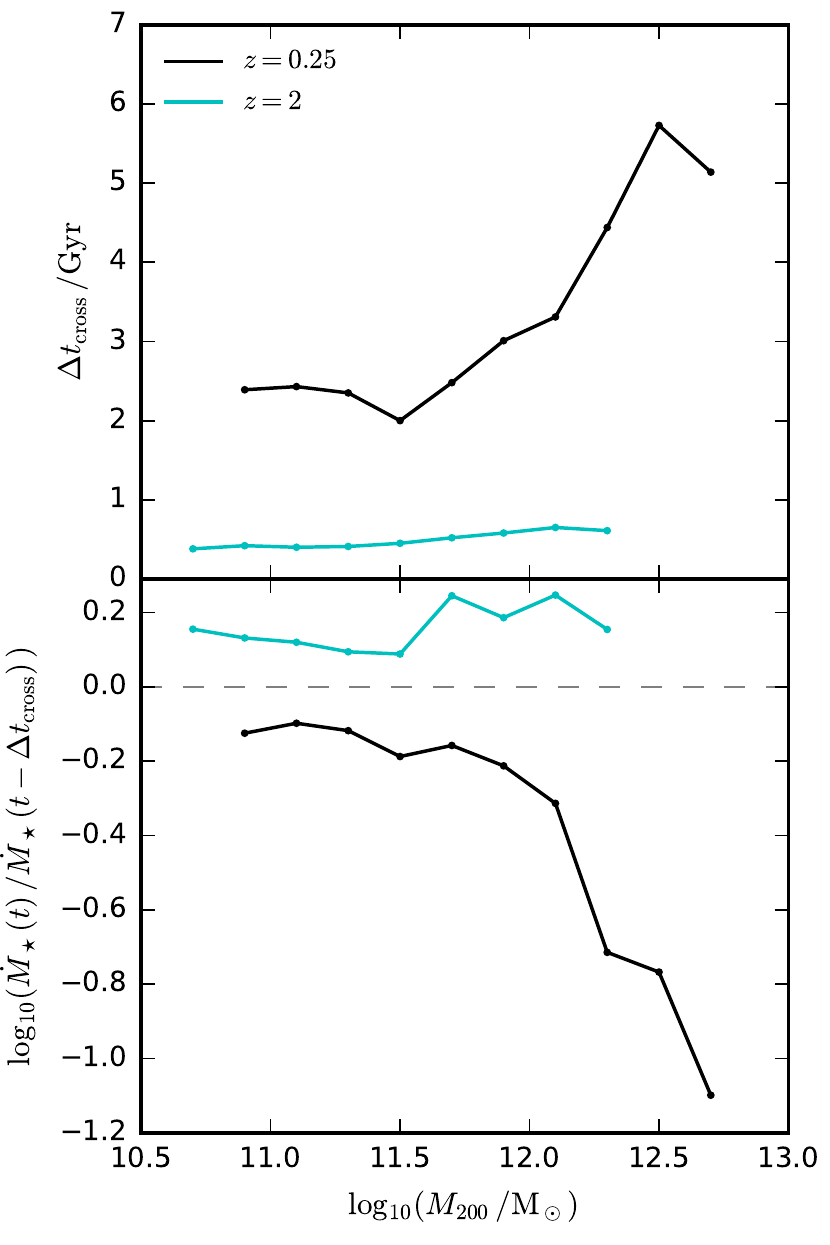}
\caption{\textbf{\textit{Top:}} the mean crossing time for outflowing particles
  ejected from the ISM to reach the virial radius (recorded at the time that
  particles reach the virial radius), plotted as a function of halo mass.
  \textbf{\textit{Bottom:}} the average change in star formation rate (compared
  to the main progenitor galaxy) computed over the crossing timescale shown
  in the top panel.
  Since galaxy star formation histories (and AGN activity) in \eagle on average peak at $z \approx 2$,
  the outflow rate at the virial radius at $z=2$ reflects in part the lower
  star formation/AGN activity in past progenitors, whereas $z=0$ outflows
  at the virial radius reflect the higher past star formation/AGN activity.
}
\label{time_delay_fig}
\end{figure}

  Another effect that turns out to be important for understanding the
  statistical trend of elevated mass fluxes at the virial radius is
  connected to the finite time taken for energy injected into the ISM
  to propagate outwards to the virial radius. Comparing the outflow
  rates at the virial radius to the rates of gas leaving the ISM
  (Fig.~\ref{mass_loading_ratio}), the ratio of the two is of
  order unity for $M_{200} \sim 10^{12} \, \mathrm{M_\odot}$
  at $z \approx 2$ but has increased by $\approx 0.5 \, \mathrm{dex}$
  by $z=0$. The top panel of Fig.~\ref{time_delay_fig} shows the
  mean crossing time for outflowing gas to move from the ISM to the
  virial radius. This time is not negligible compared to a Hubble time,
  and outflows at the virial radius will presumably at least partly
  reflect the energy injection rate at the time outflows were launched
  from the ISM.

  The bottom panel of Fig.~\ref{time_delay_fig} then shows how
  the present star formation rate (at the time the selected particles
  are leaving the halo) compares to the star formation rate in the
  past when the particles left the ISM. Due to the shape of star
  formation histories in \eagle that peak at $z \approx 2$
  \cite[see e.g. figure 9 of][]{Mitchell18}, star formation rates
  (and also AGN activity) were higher in the past for the progenitors of galaxies at $z=0$ (black line),
  but were lower in the past for the progenitors of galaxies at
  $z=2$ (cyan line). While this partially helps to explain the elevated
  outflow rates at the virial radius at $z=0$, the magnitude
  of the effect is too small to be the main explanation.
  Time delay effects do however present a convincing explanation
  for the redshift dependence of the ratio of mass loading factors
  seen in Fig.~\ref{mass_loading_ratio}. The offset between
  the star formation increase/decrease over a crossing time
  is around $0.3 \, \mathrm{dex}$ between $z=2$ and $z=0$,
  which is comparable to (and goes in the right direction to explain) the
  redshift evolution of the mass loading ratio shown in
  Fig.~\ref{mass_loading_ratio}. In addition, the change in
  star formation rate is more stark for high-mass haloes with
  $M_{200} > 10^{12} \, \mathrm{M_\odot}$ (due to a longer crossing time),
  which helps to explain the mass dependence of the mass loading
  ratio in this mass range.

  There are other factors that may contribute to the change in
  outflow rate with radius in \eagle, which we now briefly consider.
  One is that satellites may play a role by injecting energy directly
  at larger radii. We have checked this explicitly, and find that
  the energy injection rate is generally negligible compared to the
  central energy injection rate, and to the energy flux at each
  radius. % This holds up to 1e13 at least in halo mass, might be different in clusters of course
  A second physical effect that has been discussed recently within
  the context of stellar feedback-driven outflows is buoyancy
  \cite[][]{Bower17,Keller20}, which has also long-been considered
  as an important part of how AGN feedback may operate in the
  intra-cluster medium \cite[e.g.][]{Churazov02,Chandran07,Pope10}, albeit generally
  with additional physics to what is simulated in \eagle (e.g. cosmic
  rays, thermal conduction).
  Since we find in Fig.~\ref{pressure_fig} that outflows in
  $M_{200} = 10^{12} \, \mathrm{M_\odot}$ mass haloes
  are over-pressured relative to the ambient CGM, we do not
  expect buoyancy to be the main driver, as buoyancy
  becomes dynamically important as a mechanism to lift
  low-entropy gas within a multi-phase medium that is locally in
  pressure equilibrium. This situation may change in
  higher-mass haloes however ($M_{200} \sim 10^{13} \, \mathrm{M_\odot}$, not shown),
  for which winds are still over-pressured, but the
  ambient medium itself is in overall equilibrium with
  the gravitational potential.
  Finally, non-feedback related energy sources could in principle
  act on larger spatial scales to drive elevated mass fluxes
  at the virial radius. While non-trivial to check, the naive expectation
  is that gravitation-related motions would peak for gas moving
  near the halo center, where the maximum amount of potential energy
  has been converted into kinetic form. On the other hand,
  compressive heating associated with gravitational infall (and
  in particular halo mergers) could over-pressurise the CGM
  and drive large-scale outflows in the same manner as previously
  described for feedback-driven winds. From looking
  at individual outflow events in time series (not shown for conciseness),
  we find that significant outflow at the virial radius is always preceded
  by an intense but short-lived outflow event at the halo center, triggered by a period of
  star formation or AGN activity. This confirms that the large
  scale outflows are at least correlated with feedback activity, but
  on the other hand star formation and feedback will also be correlated
  with gravitational infall and mergers, so we do not draw any firm
  conclusions.

\subsection{Impact of AGN feedback}
\label{agn_impact}

\begin{figure}
\includegraphics[width=20pc]{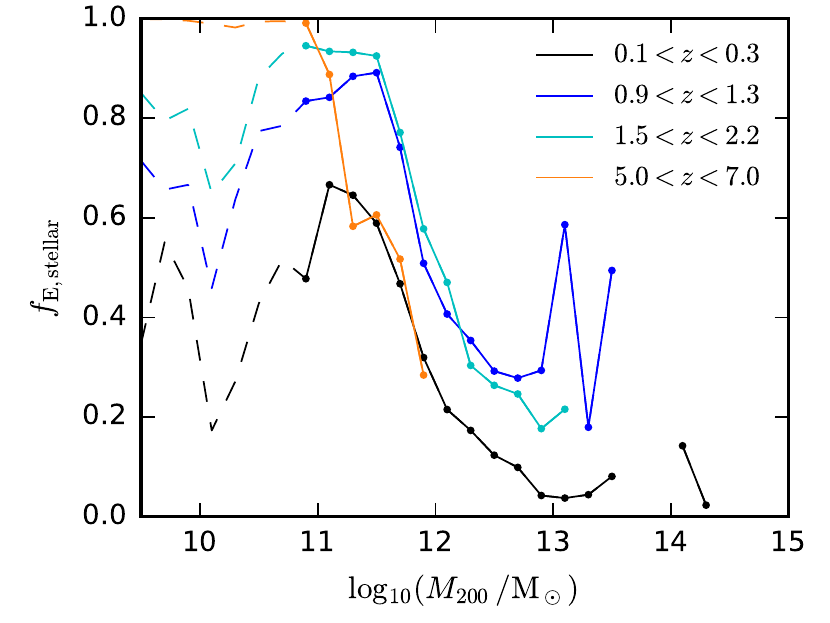}
\caption{The average fraction of energy injected by stellar feedback (as opposed to AGN feedback), plotted as a function of halo mass.
Solid (dashed) lines indicate the halo mass range within which galaxies contain on average more (fewer) than $100$ stellar particles.
Data are taken from the $50 \, \mathrm{Mpc}$ reference run, using trees with 28 snapshots.
Stellar feedback provides most of the injected energy for haloes with $M_{200} \sim 10^{11} \, \mathrm{M_\odot}$,
whereas AGN feedback dominates for haloes with $M_{200} \gtrapprox 10^{13} \, \mathrm{M_\odot}$.
The dip in the stellar feedback fraction seen at $M_{200} \sim 10^{10} \, \mathrm{M_\odot}$ is related to the halo mass at which SMBHs are seeded.
}
\label{frac_sne}
\end{figure}

Fig.~\ref{frac_sne} shows the average fraction of feedback energy injected by stellar feedback,
with the remainder contributed by AGN feedback. Generally speaking, stellar feedback is more
important in lower mass haloes and at higher redshifts. For haloes of mass, 
$M_{200} = 10^{11} \, \mathrm{M_\odot}$, the fraction of energy contributed by AGN grows from close
to zero at $z>2$ up to about $40\%$ by $z=0$. AGN provide the majority of energy injection for 
haloes more massive than $10^{12} \, \mathrm{M_\odot}$ at all redshifts recorded.

Below $z=5$, a strong feature appears
at a characteristic halo mass of $10^{10} \, \mathrm{M_\odot}$. This feature arises because of
the implementation of supermassive black hole seeding in \eagle; black hole seeds are placed
in friends-of-friends groups of that mass. The sudden increase in AGN energy at this specific mass
scale is clearly artificial, with the newly formed black hole strongly out of equilibrium with
the surrounding ISM. We have checked and verified that this feature has a negligible effect on the median stellar mass as a
function of halo mass, by comparing simulations with and without AGN feedback.
% If time check the gas fractions in the no-AGN to see if they turn over at low mass - comment here
%Another factor is that a significant fraction of the galaxies are not forming any stars (indicated
%by dashed lines) at low halo-mass, which is related to the finite resolution of the simulation, 
%and to the metallicity dependence of the star formation threshold used in \eagle.

\begin{figure}
\includegraphics[width=20pc]{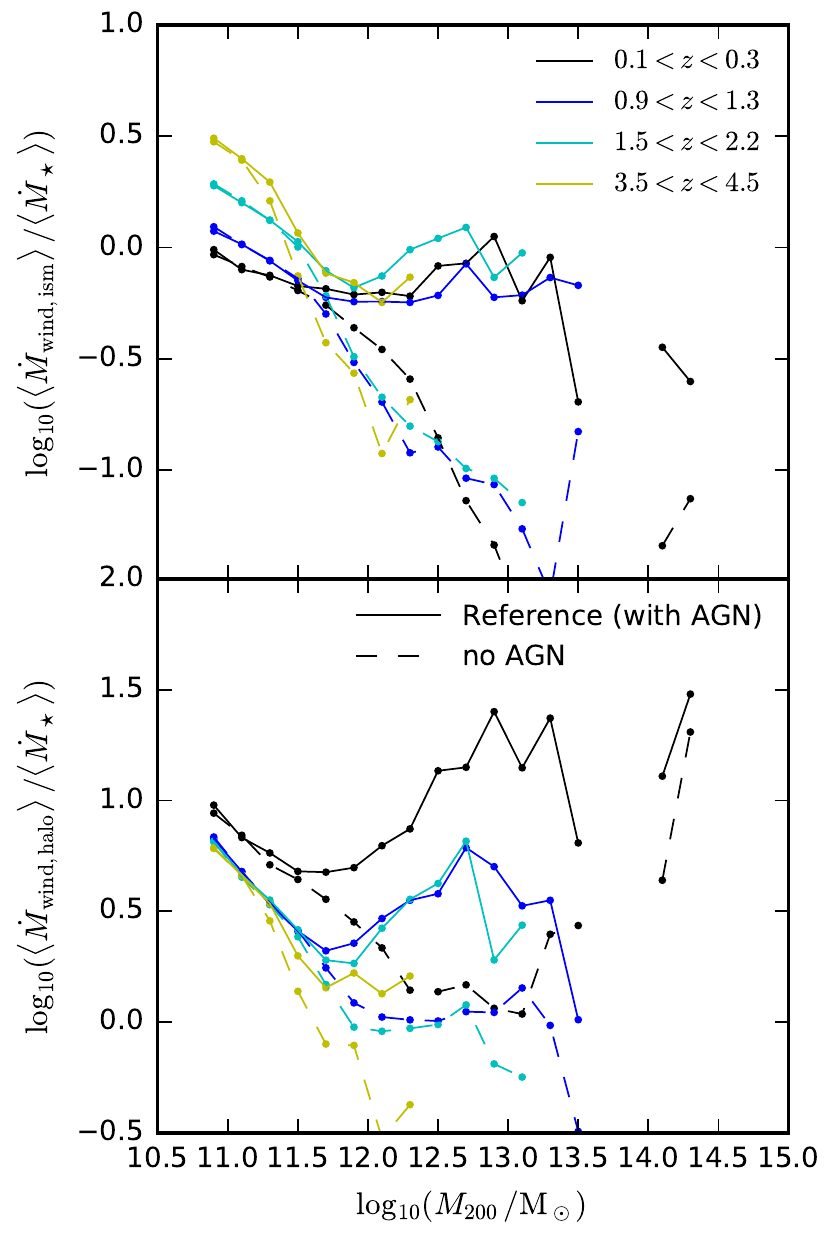}
\caption{Impact of AGN feedback on mass loading factors associated with galaxy-scale (top) and halo-scale (bottom) outflows.
Solid lines indicate outflow rates for the reference simulation (which includes AGN feedback).
Dashed lines indicate the corresponding rates for the no-AGN variant of the reference simulation.
Data are taken from the $50 \, \mathrm{Mpc}$ reference and no-AGN runs, both using trees with 28 snapshots.
Data are shown for mass bins within which galaxies contain on average more than $100$ stellar particles.
AGN feedback starts to appreciably affect outflow rates in haloes with masses $M_{200} > 10^{11.5} \, \mathrm{M_\odot}$, causing a flattening
(or upturn) of the scaling of the mass loading factor with increasing halo mass.
}
\label{outflow_agn_comp}
\end{figure}

Fig.~\ref{outflow_agn_comp} compares the outflow rates in simulations with and without AGN feedback.
We perform this comparison in terms of mass loading factors to account for the
difference in star formation activity between the two simulations at fixed halo mass.
For the galaxy-scale outflows (top panel), AGN feedback is clearly responsible for the
upturn in the mass loading factor for haloes with $M_{200} > 10^{12} \, \mathrm{M_\odot}$.
A similar picture emerges for the halo-scale outflows (bottom panel).
%It is notable that there is still a flattening of the mass loading (and a possible upturn for halo-scale outflows) for 
%the no-AGN simulation at high halo masses; we interpret this as contamination of our outflow rate measurements 
%from dynamical gas motions associated with gravitational infall, and from expansion related to gravitational heating, 
%rather than as evidence for an increase in the efficiency of stellar feedback in massive galaxies. 
% I commented this part out because with the time delay effects, this could in principle be related to stellar feedback!

\section{Numerical convergence}
\label{Convergence_sec}

As per the results and discussion presented in \cite{Schaye15}, the basic outputs of
of the \eagle simulations (e.g. the galaxy stellar mass function, see their figure 7)
are not converged with numerical resolution for a fixed set of model feedback parameters,
primarily because the anticipated radiative losses depend on the distribution
of ISM densities, which itself changes with numerical resolution.
\cite{Schaye15} argue that this convergence test (dubbed ``strong numerical
convergence'') is overly stringent
for cosmological simulations in which the ISM is
unresolved. Because the subgrid parameters of such simulations in any
case require calibration, they instead introduce the concept of ``weak numerical convergence'',
for which the change in radiative losses associated with changing resolution
is accounted for by adjusting the efficiency of feedback parameters
until agreement with the basic observables used for calibration is (re)achieved.
While clearly less demanding than a conventional (``strong'') convergence test, a weak convergence
test is still of significant utility, for example to identify the mass scales at
which non-convergence of the quenched fraction of galaxies is being driven by
sampling issues (e.g. too few star particles), rather than by purely
feedback-related issues \cite[][]{Furlong15}.

\begin{figure*}
\begin{center}
  \includegraphics[width=40pc]{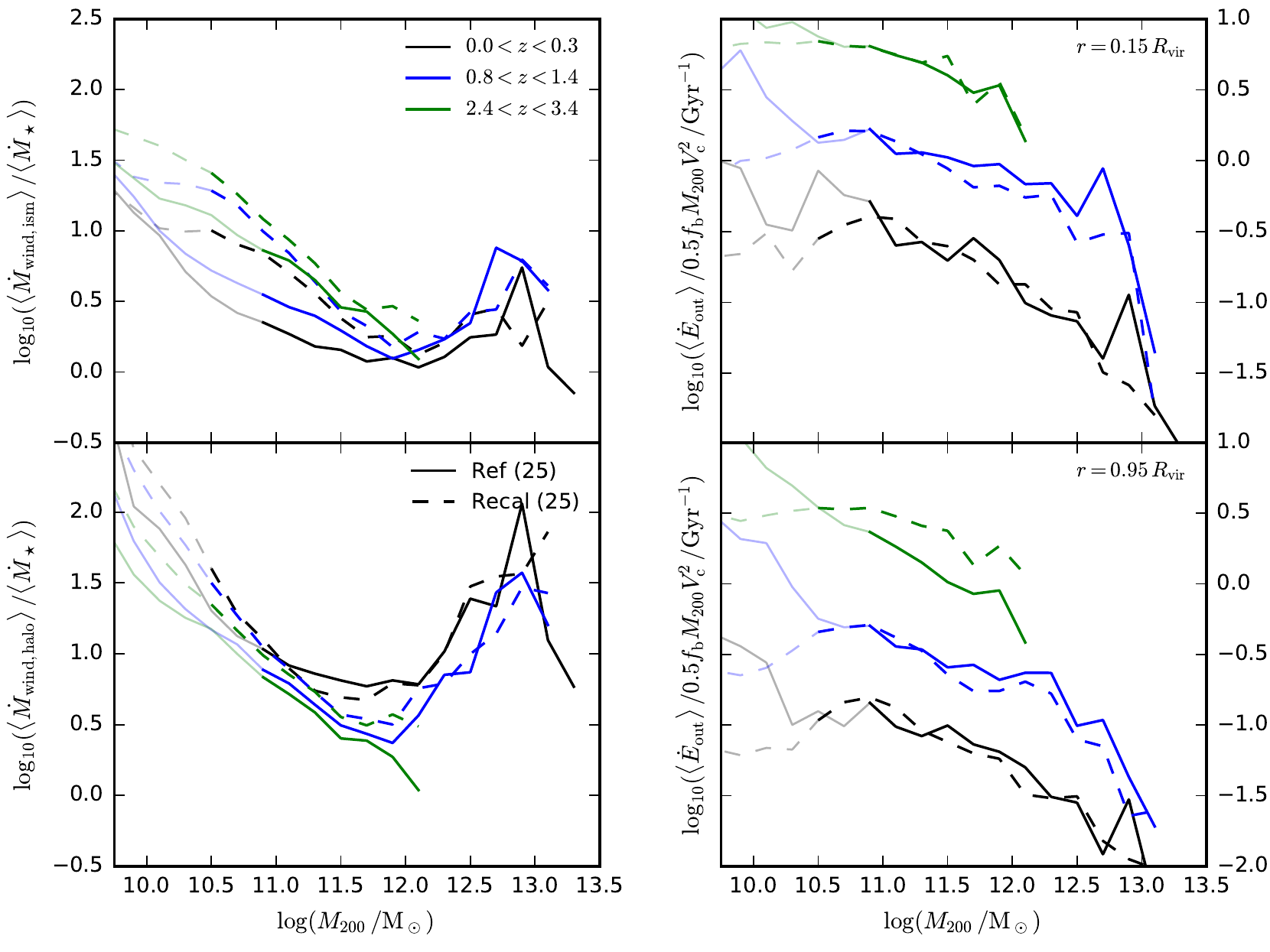}
  \caption{A comparison of outflow scalings between the reference
  \eagle $(25 \, \mathrm{Mpc})^3$ simulation (solid lines), and a higher resolution
  (eight times higher mass resolution) recalibrated simulation of the same volume (``Recal'', dashed lines).
  \textbf{\textit{Top-left:}} mass loading factors for gas leaving the ISM.
  \textbf{\textit{Bottom-left:}} mass loading factors for gas leaving the halo.
  \textbf{\textit{Top-right:}} energy flux (kinetic plus thermal, normalised by characteristic halo energy) of outflowing gas at $r=0.15 \, R_{\mathrm{vir}}$.
  \textbf{\textit{Bottom-right:}} energy flux for outflowing gas at $r=0.95 \, R_{\mathrm{vir}}$.
  Dark (light) lines indicate the halo mass range within which galaxies contain on average more (fewer) than $100$ stellar particles.
  Mass loading factors are reasonably well converged at the halo scale for $M_{200} > 10^{11} \, \mathrm{M_\odot}$, but are not quantitatively
  converged at the ISM scale for $M_{200} < 10^{12} \, \mathrm{M_\odot}$,
  in the stellar feedback regime. Energy fluxes are generally converged at both scales for galaxies with more than $100$ stellar particles.
  }
\label{recal_comp}
\end{center}
\end{figure*}

It is important then to check if outflow
rate scalings change, while still (as far as possible) retaining agreement
with the observed galaxy stellar mass function. 
Since we expect the galaxy stellar mass function to primarily reflect
the balance between gaseous inflows and outflows, the naive expectation
is that a ``weakly''-converged pair of simulations with different
resolutions (both calibrated to reproduce the observed galaxy stellar
mass function) would produce similar outflow scalings.
Fig.~\ref{recal_comp} compares the mass loading factor (left panels)
at ISM (top) and halo (bottom) scales between two \eagle simulations
with equivalent volume ($25^3 \, \mathrm{Mpc}^3$): the first with the
reference \eagle resolution and parameters, and the second with eight
times higher mass resolution and recalibrated parameters, which we refer to as the ``Recal''
simulation. Mass loading factors are significantly higher at all
redshifts in the higher-resolution Recal simulation for
$M_{200} < 10^{12} \, \mathrm{M_\odot}$ at the ISM scale,
and for $M_{200} < 10^{11} \, \mathrm{M_\odot}$ at the halo scale.
(Weak) convergence appears
better at higher masses, although statistics are too sparse to
make a robust conclusion for group and cluster-mass haloes.
Since haloes of $M_{200} \approx 10^{10.8} \, \mathrm{M_\odot}$ contain
on average only $100$ stellar particles in the reference model at
standard \eagle resolution, we conclude that there is
reasonable weak convergence for mass loading factors at the halo scale
for resolved haloes, but not at the ISM scale for
$M_{200} < 10^{12} \, \mathrm{M_\odot}$.
 
  The right panels of Fig.~\ref{recal_comp} show a
  comparison of the energy (thermal plus
  kinetic) of outflowing gas in the two simulations.
  Outflow energetics are much better converged than
  mass loading factors at the ISM scale, showing only a significant discrepancy at the
  halo scale at high redshift.
  Given that the Recal model is calibrated against
  the same observed stellar mass function as the reference model
  run at lower resolution, this implies that outflow
  energetics are a better indicator of the efficiency
  of feedback in regulating galaxy growth.
  Furthermore, since convergence is better for
  mass loading factors is better at the halo scale,
  we can also infer that galaxy formation in the
  simulation is being regulated primarily on CGM
  scales, as a consequence of the work done by energy
  injected into the CGM by feedback
  \cite[this interpretation aligns with the analysis of]
       [who find that the CGM mass fraction strongly correlates with the star formation rates in galaxies in \eagle]{Davies19}. 
  This regulation is achieved by shaping inflow
  rates of gas onto galaxies,  which in an upcoming
  study we will show are higher in
  the higher-resolution Recal simulation (both
  for recycled and first-time infalling gas),
  which explains how the simulation produces the
  same galaxy stellar mass function despite producing
  different mass loading factors at the ISM scale.

Given the recent focus with cosmological simulations on
  the question of convergence with numerical resolution in the
  CGM \cite[for column densities, ionization state, etc, ][]{VanDeVoort19,Peeples19,Suresh19,Hummels19},
  we briefly mention the possible implications of this
  for the results presented here. While something that has not been
  explicitly studied to our knowledge in cosmological simulations, it seems probable
  that outflows could be affected by CGM resolution, as this
  will (for example) affect levels of mixing with ambient gas
  via instabilities. The convergence test we present here is
  suggestive, in that we find higher inflow and outflow rates in the
  CGM for the same outflowing energy flux  (implying feedback
  is less effective at disrupting infall at higher resolution),
  but is inconclusive in that the outflowing mass fluxes also change at
  the scale of the ISM, before any interaction with the CGM
  can occur.
  
  To summarise, we find that quantitatively the mass loading
  scalings in \eagle are reasonably well converged
  at the halo scale over the mass range where galaxies are
  resolved by more than $100$ stellar particles at standard
  resolution ($M_{200} > 10^{11} \, \mathrm{M_\odot}$),
  once feedback parameters are recalibrated against observational
  constraints. Quantitative convergence is not achieved at the ISM
  scale for $M_{200} < 10^{12} \, \mathrm{M_\odot}$,
  but qualitatively the picture for outflows in \eagle remains
  the same at the higher resolution explored:
  the mass loading factor scales strongly
  with halo mass, with a minimum value at
  $M_{200} \sim 10^{12} \, \mathrm{M_\odot}$, and outflow rates
  are elevated at the virial radius compared to at
  the boundary of the ISM, especially at low redshift.

\section{Literature comparison}
\label{discussion_section}

Here, we conclude our analysis of outflows
by comparing to a range of models, simulations and observations from the literature, and 
explore the conclusions that can be drawn from this wider context.

\subsection{Comparison to semi-analytic models}

\begin{figure}
\includegraphics[width=20pc]{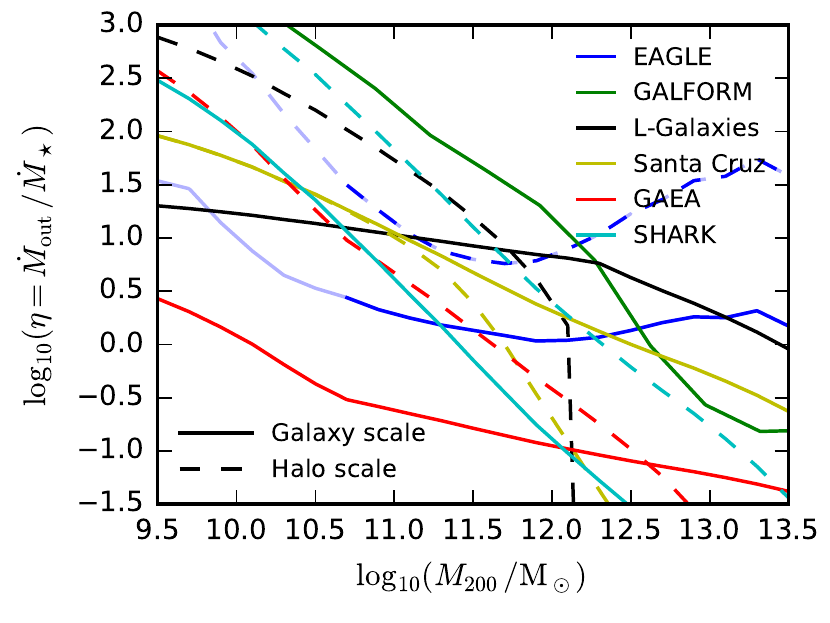}
\caption{
A comparison of mass loading factors between \eagle and a set of semi-analytic models from the literature, plotted as a function of halo mass at $z=0$.
Semi-analytic models shown include specific implementations of the GALFORM \protect{\citep{Mitchell18}}, L-Galaxies \protect{\citep{Henriques15}}, 
Santa Cruz \protect{\citep{Somerville15}}, SHARK \protect \citep{Lagos18}, and GAEA \protect{\citep{Hirschmann16}} models.
Outflow rates are plotted for gas being ejected from the ISM (solid lines), and for gas being ejected from the halo virial radius (dashed lines).
Each line colour corresponds to a given model, as labelled.
For \eagle, dark (light) lines indicate bins within which galaxies contain on average more (fewer) than $100$ stellar particles.
All of the semi-analytic models shown (and \eagle) are tuned to match the local galaxy luminosity function and/or the galaxy stellar mass function.
This agreement can apparently be achieved with wildly different scenarios for how much gas outflows from the ISM, and from the halo, emphasising the deeply
degenerate nature of galaxy evolution if only stellar mass constraints are considered.
%\eagle data are taken from the $100 \, \mathrm{Mpc}$ reference run, using trees with 200 snipshots.
}
\label{sam_comparison}
\end{figure}

Semi-analytic models are an established method to study the evolution of galaxies within
the full cosmological context \cite[see][for an overview]{Baugh06,Somerville14}. Most
semi-analytic models assume that stellar feedback drives galactic outflows from the ISM
of galaxies, with a mass loading factor that scales negatively with galaxy circular 
velocity \cite[e.g.][]{Kauffmann93,Cole00}. This in turn allows the models to achieve
a match with the faint end of the galaxy luminosity function \cite[e.g.][]{Benson03}
\footnote{There are alternative pictures that have been considered, such as the pre-heating scenario explored for example in \cite[][]{Lu15}.}.
Our measurements of outflow rates from \eagle are (deliberately) suitable for direct
comparison to the prescriptions assumed in semi-analytic models, and we show a direct
comparison to a subset of recent models from the literature in Fig.~\ref{sam_comparison}.

It is immediately apparent from Fig.~\ref{sam_comparison} that there is an enormous 
dispersion in what is assumed for the mass loading factor from one model to another 
(up to nearly four orders of magnitude at a given halo mass), despite the fact that
all the models shown are calibrated to reproduce the observed distribution of stellar mass.
Focussing only on the normalisation, the large differences in mass loading factor are driven
by two factors. First, each model makes different assumptions regarding the level of dichotomy
between outflow rates of gas leaving the ISM (solid lines) versus the halo virial radius 
(dashed lines). The \cite{Henriques15}, \cite{Hirschmann16} and \cite{Lagos18} models \cite[all at least partially adapted from the 
L-galaxies model of][]{Guo11} prescribe the excess energy remaining in galactic
winds after they have escaped the ISM, and assume this energy can drive even greater
amounts of gas out of the halo. Conversely, the \galform and Santa Cruz models
assume that the amount of gas ejected from the halo is equivalent (or less than for
the Santa Cruz model) to the amount of gas ejected from the ISM 
\cite[e.g.][]{Somerville08, Mitchell18}. Both scenarios are degenerate in terms of
stellar mass assembly, in the sense that they both reduce the fraction of baryons that form stars.

The second explanation for the differences in mass loading normalisation stems from the assumed
efficiency of recycling of ejected wind material. For example, the \galform model
assumes a very efficient recycling timescale that is of order the halo dynamical time (such that ejected gas returns in only $10\%$ of a Hubble time),
whereas the Santa Cruz model assumes that gas returns over a Hubble time. This forces
the former model to invoke mass loading factors that are much larger than the latter.
Again, these scenarios are degenerate in terms of stellar mass assembly \cite[e.g.][]{Mitchell14}, 
at least up until the point that the recycling timescale becomes so long that galaxy clusters 
no longer retain the universal baryon fraction \cite[][]{Somerville08}.

% Leave this for another day
%Similar to the differences in normalisation, differences in the slope of the scaling
%between mass loading and halo mass can coexist by introducing a negative scaling between the 
%efficiency of ejected gas recycling and halo mass, which permits a shallower negative
%scaling for the mass loading factor while still reproducing the observed faint-end luminosity 
%function slope \cite[e.g.][]{Henriques13}.

Given this (long-standing) impasse, it is then interesting to consider the picture emerging from modern
hydrodynamical simulations. The full simulation picture is shown in Section~\ref{sim_comp_sec}, but
we choose to show the direct comparison between semi-analytic models and \eagle here.
The outflow rates from \eagle (blue lines) are qualitatively
closer to the scenarios presented by the GAEA \cite[red lines, ][]{Hirschmann16},
SHARK \cite[cyan lines, ][]{Lagos18}, and L-galaxies \cite[black lines, ][]{Henriques15} models, in that significantly more
gas is ejected from halo virial radii than from the ISM. Quantitatively however, \eagle differs significantly
in both normalisation and slope with the L-galaxies model shown.
\cite{Hirschmann16} adopt a mass loading prescription for
gas leaving the ISM inspired by the \fire simulations \cite[][]{Hopkins14}, as measured by \cite{Muratov15}.
Qualitatively, the picture from this model is close to that seen in \eagle at $z=0$, with a relatively low
normalisation and fairly shallow scaling of the galaxy-scale mass loading factor, combined with
a significantly higher normalisation for the outflow rates at the halo virial radius.
We present a direct comparison with \fire and other hydrodynamical simulations in the following
section.

Finally, we note that the mass loading factors shown for the semi-analytic models are
for stellar feedback only. The upturn in mass loading factors for high-mass galaxies in \eagle is caused 
by AGN feedback. Most semi-analytic models assume that AGN feedback acts only to suppress inflows 
rather than drive AGN outflows directly\footnote{The exception for the models shown here is 
\cite{Somerville08}, which does include AGN-driven outflows from the ISM. We cannot however easily
infer outflow rates at a given halo mass from their prescription for AGN feedback, so we show their prescription for 
stellar feedback only.}, which is qualitatively different from the scenario presented in \eagle.
We note that semi-analytic models where AGN do eject baryons from haloes have been considered as an explanation
for the observed X-ray luminosity of galaxy groups \cite[][]{Bower08,Bower12}.

\subsection{Comparison to other cosmological simulations}
\label{sim_comp_sec}

% Note should be doing stellar mass in 30kpc apertures for Eagle to compare to Illustris here I think.
\begin{figure*}
\begin{center}
\includegraphics[width=40pc]{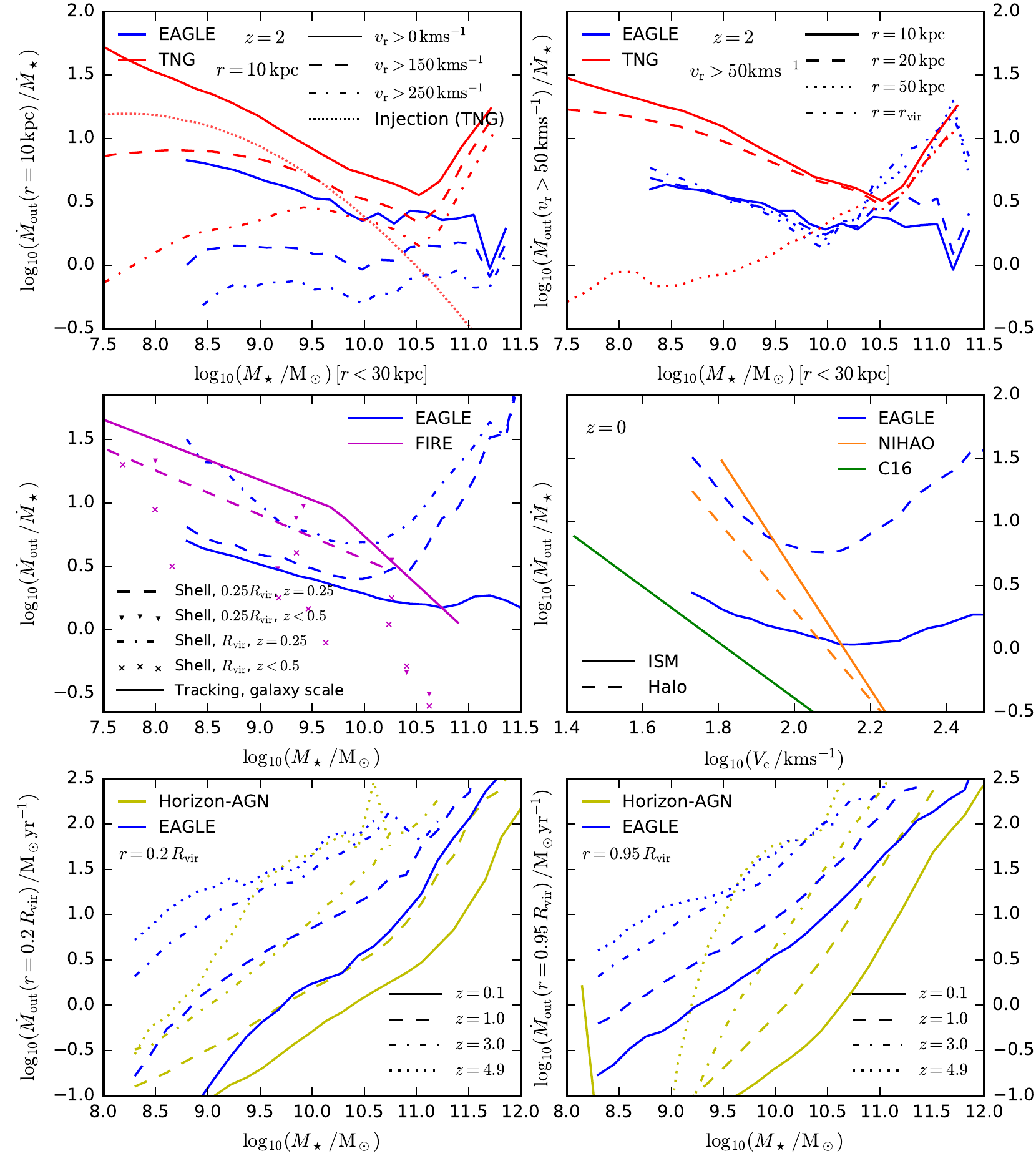}
\caption{Comparison of wind mass loading factors and outflow rates between \eagle and other recent hydrodynamical simulations from the literature.
%All \eagle data are taken from the $100 \, \mathrm{Mpc}$ reference simulation, using trees constructed with 200 snipshots.
\textbf{\textit{Top-left:}} compares \eagle (blue) and Illustris-TNG \protect{\citep[red]{Nelson19}} at $z=2$, showing the median mass loading factor for gas at $r=10 \, \mathrm{kpc}$, 
plotted as a function of stellar mass.
Different line styles indicate different minimum radial velocity cuts, as labelled.
For Illustris-TNG, we also show the mean mass loading factor applied at injection (dotted red line).
\textbf{\textit{Top-Right:}} mass loading for gas at different distances from the halo center, as labelled.
In this case gas is selected with radial velocity $v_{\mathrm{r}} > 50 \, \mathrm{km s^{-1}}$.
For \eagle, we only show measurements at $50 \, \mathrm{kpc}$ for galaxies with $M_\star > 10^{9} \, \mathrm{M_\odot}$, below which gas at $50\,\mathrm{kpc}$ is outside the halo virial radius 
(we instead show measurements for a shell at the virial radius with the blue dash-dotted line).
\textbf{\textit{Middle-left:}} compares \eagle and the \fire zoom-in simulations (magenta). Note that the FIRE simulations do not include AGN feedback.
Dashed blue and magenta lines compare shell-based measurements of the mass loading at $r=0.25 R_{\mathrm{vir}}$, at redshift $z=0.25$ \protect{\citep{Muratov15}}.
For \eagle, the dashed-dotted blue line shows the same but for a shell at the virial radius.
For \fire, individual galaxies are shown by the magenta points for shells at different radii, as labelled \protect{\citep{Muratov15,Muratov17}}. 
Solid lines show (tracking-based) mass loading factors for gas being ejected from the ISM, time-integrated over the entire history of each galaxy \protect{\citep{AnglesAlcazar17}}.
%These time-integrated measurements are presented for galaxies at $z=0$ in \eagle, and for a range of redshifts in \fire.
\textbf{\textit{Middle-right:}} compares \eagle, NIHAO (orange), and the simulations presented in \protect \cite{Christensen16} (green) at $z=0$, 
showing (tracking-based) mass loading factors for gas being ejected from the ISM of galaxies (solid lines), plotted as a function of halo circular velocity at the virial radius.
For \eagle and NIHAO, we also show mass loading factors for gas being ejected through the halo virial radius (dashed lines).
Note that only \eagle includes AGN feedback.
\textbf{\textit{Bottom:}} compares \eagle and the Horizon-AGN simulation \protect{\citep[yellow]{Beckmann17}}, showing outflow rate as a function of stellar mass.
Outflow rates are computed with a shell-based method at $r=0.2 \, R_{\mathrm{vir}}$ (bottom-left), and at $r=0.95 \, R_{\mathrm{vir}}$ (bottom-right).
}
\label{sim_comparison}
\end{center}
\end{figure*}

Fig.~\ref{sim_comparison} presents an overview of the mass loading factors from recent cosmological hydrodynamical
simulations. Each study shown uses a different method to measure outflow rates, and we have taken
care to (as far as is reasonably possible) compare \eagle to other simulations using equivalent measurements.

The upper panels of Fig.~\ref{sim_comparison} compare \eagle to the $50^3 \, \mathrm{Mpc}^3$ Illustris-TNG (TNG-50) simulation at $z=2$,
taking measurements from \cite{Nelson19}. \cite{Nelson19} measure outflow rates in shells at a given physical distance from the halo
center, for gas radially outflowing faster than some minimum radial velocity cut (different line styles in the upper-left panel show
different cuts). These simple criteria are straightforward to implement, and so we can perform a like-for-like comparison of the
simulations at $z=2$ \cite[the redshift focussed on by][]{Nelson19}. Taking all outflowing gas with
$v_{\mathrm{r}}>0 \, \mathrm{km s^{-1}}$ at a distance of $10 \, \mathrm{kpc}$ (solid lines in the top-left panel), \eagle and
TNG-50 display qualitatively similar behaviour for stellar masses, $M_\star < 10^{10.5} \mathrm{M_\odot}$, but are offset in
normalisation by up to $0.5 \, \mathrm{dex}$, with higher mass loading factors in TNG-50 than in \eagle.

% TNG recoupling is done at density = 0.05 times the density threshold for star formation
Mass loading for stellar feedback is set by hand at injection for TNG-50 (shown as the dotted red line), with
outflows seeded by wind particles that are decoupled from the hydrodynamical 
scheme until they reach a density below $n_{\mathrm{H}} \sim 0.005 \, \mathrm{cm^{-3}}$ \cite[][]{Pillepich18}.
In practice the TNG outflows generally recouple within $10 \, \mathrm{kpc}$, and the mass loading factors compared
with \eagle here were measured on scales at which the direct contribution of decoupled particles to the outflow
is negligible \cite[][]{Nelson19}. In addition, outflows at these scales in TNG may have to do significant work against the magnetic
pressure of circum-galactic gas, which is not accounted for in \eagle.
The TNG mass loading at injection (minus a residual metallicity dependence) is set to scale negatively with circular
velocity as $V_{\mathrm{c}}^{-2}$. Although the measured outflow rate is slightly higher than the injected one,
they track each other closely at low mass, where stellar feedback dominates over AGN feedback \cite[][]{Nelson19}.
No mass loading factor is imposed by hand in \eagle, but the feedback model is still calibrated against
  similar observational constraints to those used for TNG, and so it is encouraging (but not surprising) to see that the mass
  scaling of the mass loading factor is similar between the simulations in the stellar feedback-dominated regime.

At higher stellar masses, \cite{Nelson19} report a strong upturn in the mass loading factor that is attributed to AGN feedback.
A weaker upturn for galaxy-scale outflows at $10^{13} \, \mathrm{M_\odot}$ haloes is seen in \eagle in Fig.~\ref{mass_loading}, but is not visible
using the shell-based measurements at $10 \, \mathrm{kpc}$, where the mass loading instead flattens at high stellar masses.
The upper-right panel of Fig.~\ref{sim_comparison} compares shell-based outflows at different radii, and here a clear upturn in the mass loading
is visible in \eagle at a distance of $50 \, \mathrm{kpc}$ from the halo center (dotted blue line), similar to that seen in TNG-50 at all radii.
This indicates a significant difference in the smaller-scale wind launching for AGN feedback between the simulations, with TNG-50 ejecting
large amounts of gas from the center of massive galaxies, while \eagle launches relatively little gas but with the wind seemingly continuing
to load mass as a function of radius, such that the mass loading increases out to the virial radius (dash-dotted blue line).

\interfootnotelinepenalty=10000
Comparing the mass loading in the stellar feedback regime in the upper-right panel of Fig.~\ref{sim_comparison} reveals further stark differences
between the two simulations. While TNG-50 ejects significantly more gas per unit star formation than \eagle at $10 \, \mathrm{kpc}$ in
low-mass galaxies, the outflows seem to decline strongly as a function of radius in TNG-50. Outflows behave differently in \eagle,
with mass loading that either stays roughly constant with, or grows with, radius. As such, the mass loading factor at $50 \, \mathrm{kpc}$
is about $0.5 \, \mathrm{dex}$ higher in \eagle for galaxies of stellar mass $M_\star \sim 10^9 \, \mathrm{M_\odot}$\footnote{Note that for convenience we do not show
mass loading factors at a distance of $50 \, \mathrm{kpc}$ for galaxies with stellar masses below $10^9 \, \mathrm{M_\odot}$ in \eagle. This typically
selects gas outside the halo virial radius, where we cannot make measurements without incurring significant additional computational cost to associate
particles with haloes.}.
This difference implies that there is likely a large difference in the efficiency of recycling of ejected wind material between \eagle and TNG-50
(with recycling being a more important source of inflows in TNG-50 than in \eagle),
which presumably affects the observable properties of the circum-galactic medium as a function of impact parameter from galaxies.
\cite{Davies20} find a very consistent picture by comparing the total baryon content of haloes between \eagle and Illustris-TNG, which
they show is much higher in TNG than in \eagle at low mass.

The middle-left panel of Fig.~\ref{sim_comparison} compares outflow rates in \eagle with the \fire zoom-in simulations \cite[introduced in ][]{Hopkins14}.
Relative to \eagle, the \fire simulations employ significantly higher mass and spatial resolution (with the improvement scaling negatively with the mass
of the targeted haloes), allow a cold ISM phase to form without imposing a temperature floor, and implement a more explicit representation of stellar feedback 
(separating contributions from radiation, stellar winds, and type II supernova explosions). The \fire simulations do not include AGN feedback.
% Do I want to comment on the difference in SNe implementation here? Probably not for conciseness reasons. Type II supernovae are implemented using a 
We show the best-fit relation to the \fire simulations at $z=0.25$ from \cite{Muratov15}, measured using shells at one quarter of the halo virial radius
(dashed magenta line). Mimicking this type of measurement in \eagle (dashed blue line), the two simulation sets are similar but are
offset by a factor of two up until the halo mass scale ($M_{200} \sim 10^{12} \, \mathrm{M_\odot}$) where AGN feedback causes an upturn at high masses in \eagle.
We note that if the comparison is instead performed as a function of halo mass (shown in Appendix~\ref{ap_fire_comp}), the two mass loading factors agree almost perfectly between
the two simulations over the common mass range between the simulations, which can be explained if the median stellar mass at fixed halo mass is higher in \fire than in \eagle.
\cite{AnglesAlcazar17} present a complementary measurement to \cite{Muratov15} using Lagrangian particle tracking to measure particles ejected from the ISM 
\cite[solid magenta line, taken from the fit presented in][]{Dave19},
similar to our preferred methodology in this study. These measurements are presented as a cumulative integration over all outflow and star formation events
over the entire history of each galaxy shown. We perform an equivalent integration for our particle tracking-based outflow rates in \eagle, presented as the solid
blue line in the middle-left panel of Fig.~\ref{sim_comparison}. We note that both the galaxy and halo-scale outflow selection criteria differ between the two studies, although
they are both designed to in principle measure the same thing (the outflow rates of gas being ejected from the ISM/halo by feedback).

As with the comparison to TNG-50, larger differences become apparent when considering the change in the mass loading as a function of radius.
\cite{Muratov17} present measurements of the mass loading in \fire at the virial radius (magenta crosses), which can be compared to measurements
at a quarter of the virial radius (magenta triangles, or the dashed line) from \cite{Muratov15}. In most cases the mass loading is smaller at larger
radii in \fire, whereas the opposite is true in \eagle at low redshift. As with the comparison to TNG-50, this implies that recycling of
gas ejected from galaxies is likely much more efficient in \fire than in \eagle.

The middle-right panel of Fig.~\ref{sim_comparison} presents a comparison with two additional sets of zoom-in simulations,
including the simulations of \citet[][green line]{Christensen16}, and measurements of the NIHAO simulations presented by \citet[][orange lines]{Tollet19},
both as a function of stellar mass. Neither of these simulations include AGN feedback.
Both these studies utilise particle tracking-based measurements of outflows, which we compare to our particle tracking
measurements at $z=0$. \cite{Tollet19} find substantially higher mass loading factors for gas ejected from the ISM (solid orange line)
than in \eagle (solid blue line), with a very steep dependence on mass. They also find that less mass is (on average) ejected
from the virial radius (dashed orange line), which is in strong disagreement with the $z=0$ measurements from \eagle (dashed
blue line). \cite{Christensen16} find somewhat lower mass loading factors for gas being ejected from the ISM, and with
a slightly steeper mass dependence than in \eagle. They find that a substantial fraction of this gas is then ejected from the virial
radius, but do not present measurements of gas being ejected from haloes that was not previously in the ISM, making it unclear
how their simulations compare in terms of outflows at the virial radius. 

Finally, the bottom panels of Fig.~\ref{sim_comparison} present a comparison with the Horizon-AGN simulation \cite[][]{Dubois14},
showing measurements presented in \cite{Beckmann17}\footnote{We only present measurements here for their fiducial simulation
that includes AGN feedback.}. 
\cite{Beckmann17} measure outflow rates for two $2 \, \mathrm{kpc}$ thick shells at $20 \, \%$ (bottom-left)
and $95 \, \%$ of the halo virial radius, and we plot their measurements as a function of stellar mass (without any aperture
correction). 
Reproducing these measurements in \eagle, the comparison shows that outflow rates at a given stellar mass are (in most situations)
significantly higher than in Horizon-AGN (for example by about $0.5 \, \mathrm{dex}$ at $0.2 R_{\mathrm{vir}}$ at $z=0$).
We note that there are substantial differences between the low redshift galaxy stellar mass function in Horizon-AGN and \eagle, with Horizon-AGN significantly
over-predicting the stellar masses of low-mass galaxies, and \eagle under-predicting the abundance of galaxies at the knee of the mass function
\cite[][]{Schaye15,Kaviraj17}. As such the comparison performed here will be comparing galaxies hosted by dark matter haloes of differing mass.

Also of interest is the comparison between outflow rates at $0.2$ versus $0.95 R_{\mathrm{vir}}$.
For massive galaxies, both simulations eject similar or greater amounts of gas from haloes than 
through the inner surface at $0.2 R_{\mathrm{vir}}$. For lower-mass galaxies, \eagle continues
to eject similar or greater amounts through the outer surface, whereas Horizon-AGN ejects very
little gas through the virial radius compared to the inner surface.
This underlines the importance of considering outflowing flux 
as a function of scale out into the halo.

Taken at face value, the comparisons shown in Fig.~\ref{sim_comparison} indicate that hydrodynamical simulations
are seemingly able to reproduce observed stellar masses with different scenarios for gaseous outflows, similar
to the situation seen for semi-analytic models in Fig.~\ref{sam_comparison}. 
That said, while we have emphasised the differences it is also important to emphasise that there is qualitative agreement
between simulations, in the sense that all predict declining mass loading factors as a function of galaxy
mass up to $M_{200} \sim 10^{12} \, \mathrm{M_\odot}$, and are in a similar level of qualitative agreement at higher masses if AGN
feedback is included.
We caution furthermore that some of the differences between the relations shown in this
figure will arise from differences in the selection of outflowing particles (this only applies to the Lagrangian measurements), 
and so the discrepancies could be exaggerated in some cases.
In addition, the level of agreement with the observed galaxy stellar mass function is unknown for zoom-in simulations
(that must instead rely on comparison to the inferred median relationship between stellar mass and halo mass for central galaxies), and
large differences in the stellar mass function could exist between some of the different simulations shown (this is definitely
the case for Horizon-AGN).

As for the question of why galactic winds in the \eagle simulations appear to entrain more circum-galactic
gas at larger radii compared to other simulations (at least for those where such a comparison is currently
possible), we speculate that is related to the high heating temperatures adopted in the \eagle feedback model.
In reality, energy from feedback is initially injected into a far smaller mass of material compared to the
mass that is heated or kicked for the implementations of subgrid feedback models used in all cosmological
simulations, such that gas around stars and black holes will (at least locally) achieve much larger
velocities and temperatures. The choice made in \eagle to heat relative few particles to a high temperature
was motivated by this realisation, and could plausibly lead to outflows escaping the ISM with higher specific
energy than in other simulations, allowing the winds to have a greater impact on the ambient circum-galactic
medium. Explicit comparison of the energetics of outflows at different spatial scales between different
simulations would show whether or not this is indeed the case.

% How does the mechanical feedback scheme work?
% The idea is that in the snowplow phase, you inject a fixed amount of momentum into the neighbouring cells (fine).
% So indeed the specific energy of outflowing particles/cells still depends on their mass. 

\subsection{Comparison to observations}
\label{obs_comp_sec}

As discussed in the introduction to this work, our priority in this study is to measure the flows of gas leaving
galaxies and haloes using the information available from our simulations, independent of observational considerations.
Our Lagrangian methodology for selecting outflowing gas does not naturally map onto the way outflowing gas is detected
in observations, and in addition we do not explore any phase decomposition of outflowing gas.
With these caveats in mind, it is nonetheless interesting to perform a rudimentary comparison between the outflow
rates in \eagle and a best-guess for the outflow rates of real galaxies from observations.

We choose to compare to \emph{down the barrel} observations of 7 local galaxies from \cite{Chisholm17}, which use the HST-COS spectrograph to detect
multiple ultraviolet (UV) metal ions in absorption against the continuum of the associated galaxy, enabling (via
photo-ionization modelling as a function of velocity) a robust determination of the ionization structure of the outflowing
gas, in turn permitting a determination of the outflow rate. These observations are estimated to probe outflowing gas
at small scales with respect to the galaxy; \cite{Chisholm16} estimate the detected outflowing gas is within $300 \, \mathrm{pc}$
from the galaxy along the line of site. 

\begin{figure}
\includegraphics[width=20pc]{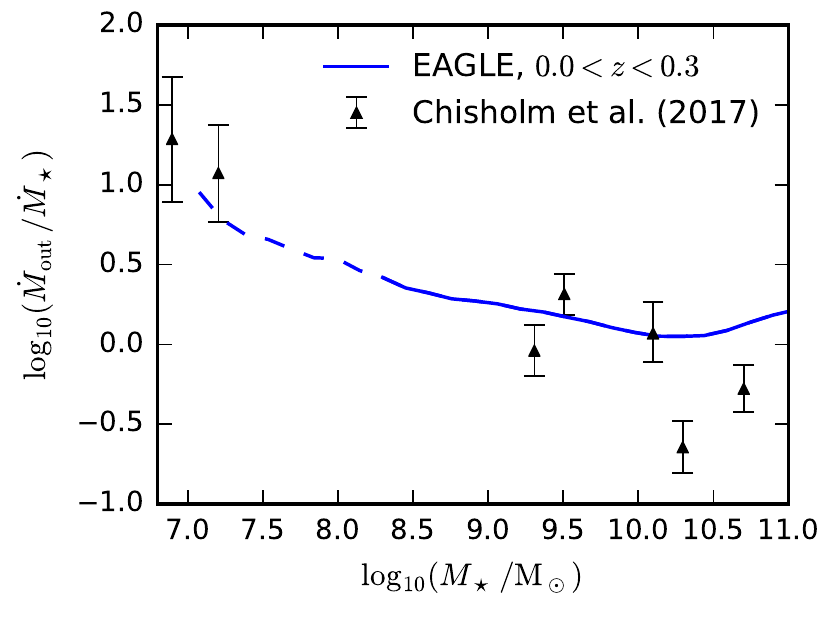}
\caption{
Comparison of mass loading factors between the \eagle simulations and estimates from ``down-the-barrel'' observations of local galaxies from \protect \cite{Chisholm17}, plotted as a function of galaxy stellar mass.
Mass loading factors from \eagle are measured for gas being ejected from the interstellar medium over a redshift range, $0<z<0.3$.
Solid (dashed) lines indicate the stellar mass range within which galaxies contain on average more (fewer) than $100$ stellar particles.
\eagle is consistent with these observations, although the level of agreement might be fortuitous,
given that we are not comparing like-for-like quantities.
%\eagle data are taken from the $100 \, \mathrm{Mpc}$ reference simulation, using merger trees with 200 snipshots.
}
\label{obs_comp}
\end{figure}

The \eagle simulations do not include all of the relevant physics (for example photo-ionization from
local radiation sources) and do not reach the resolution required to robustly mimic such a selection of gas.
Our measurements of how much gas is removed from the ISM by feedback is however still conceptually the
  same quantity as is being inferred from the observations, and so we do nonetheless present a comparison to the
  outflow rates of gas being ejected from the ISM in \eagle, shown in Fig.~\ref{obs_comp}.
The observations of \cite{Chisholm17} probe star forming galaxies in the stellar mass range where stellar feedback is expected
to dominate. They find evidence for an anti-correlation between mass loading factor and stellar mass, with a power law slope
of $-1.6$ when plotted as a function of circular velocity. Their relation is consistent with our measurements from the \eagle
simulations; we find a best-fit slope of $-1.5$ as a function of halo circular velocity, $V_{\mathrm{c}} \equiv \sqrt{G M_{200} / R_{\mathrm{vir}}}$, for
low-mass haloes ($M_{200} < 10^{12} \, \mathrm{M_\odot}$) at $z=0$.
This agreement is encouraging, and demonstrates that the outflow rates in \eagle are not implausible given current constraints.
At the same time, the level of the quantitative agreement is likely fortuitous to some extent, as we are not comparing like-for-like quantities.

% For example,
%\cite{Chisholm17} compare to the \cite{Muratov15} relationship and find they are a factor of five below, speculating that
%they may be missing a contribution from gas in molecular or hotter phases than are traced by their observations.
%Our comparison to the \cite{Muratov15} relation (middle-left panel of Fig.~\ref{sim_comparison}) reveals that \eagle is fairly consistent with the \fire simulations at 
%$z=0.25$ when the comparison is made for gas in a shell at one quarter of the halo virial radius (at least up until the mass scale
%where AGN feedback starts to play a role in \eagle), emphasising that disagreements at the level of a few can easily be explained
%away on the grounds of how measurements are made, even for simulations.
%On the other hand, models that use mass loading factors that are much larger (for example the \galform model shown in Fig.~\ref{sam_comparison})
%do start to look implausible when compared to observational constraints.

\section{Summary}
\label{summary_section}

We have presented measurements of outflow rates of gas from galaxies and from their associated 
dark matter haloes, taken from the reference \eagle hydrodynamical simulation.
We find that galactic winds are driven from the ISM in \eagle with a mass loading
factor ($\eta \equiv \dot{M}_{\mathrm{out}}/\dot{M}_\star$) that scales approximately as $\eta \propto M_{200}^{-0.5} \propto
V_{\mathrm{c}}^{-3/2}$ for low-mass galaxies ($M_{200} < 10^{12} \, \mathrm{M_\odot}$, Fig.~\ref{mass_loading}).
For reference, $\eta \propto M_{200}^{-1}$ would be required to explain the empirically inferred scaling of stellar
mass with halo mass for $M_{200} < 10^{12} \, \mathrm{M_\odot}$ using galaxy-scale outflows alone
(see discussion in Section~\ref{results_section}), implying that additional sources of mass scaling
are required to explain the agreement between \eagle and the observed galaxy stellar mass function.
We do find a scaling close to $\eta \propto M_{200}^{-1}$ when measuring outflow rates at the virial radius, 
but a discussion of the complete picture is deferred to a future study where we will present measurements
of gaseous inflow rates at different spatial scales.
Parametric fits to the mass loading factor as a function of redshift and halo mass (as well as stellar mass and halo maximum circular velocity)
are provided in Appendix~\ref{ap_parafit}.

Similar to the result found in the recent analysis of the
TNG-50 simulation of \cite{Nelson19}, we find that AGN feedback causes the 
scaling of the wind mass loading factor with mass to flatten and then increase for galaxies above a characteristic halo
mass of $10^{12} \, \mathrm{M_\odot}$ (Fig.~\ref{outflow_agn_comp}). 

We find that the mass loading factor has a steeper dependence on halo mass when
measured at the halo virial radius, and with a much clearer upturn due to AGN feedback at high masses (Fig.~\ref{mass_loading}).
We also find typically that significantly more baryons are ejected through the virial
radius than out of the ISM, particularly at low and high halo masses, and at low
redshift (Fig.~\ref{mass_loading_ratio}).
We explore a number of simplified explanations for this effect.
  We find that winds are over-pressured relative
  to the ambient CGM in \eagle, consistent with an energy-driven scenario
  in which outflows generate momentum by doing $P \mathrm{d}V$ work (Fig.~\ref{pressure_fig}).
  We also find that time delay effects play a role, as outflows at the halo
  virial radius will reflect energy injection rates in the past, which for example were
  higher for low-redshift galaxies (Fig.~\ref{time_delay_fig}).

Aside from mass fluxes, we also consider a number of outflow properties in
the simulations. We estimate that winds in \eagle typically retain $\approx 20\%$ of the energy injected
by feedback, modulated by trends with both halo mass and redshift (Fig.~\ref{energy_momentum}).
Outflow velocities cover a wide range at a given halo mass/redshift, and increase positively with redshift
and halo mass up to $M_{200} \sim 10^{12} \, \mathrm{M_\odot}$ (Fig.~\ref{velocity_ism_wind}). Below this mass the median outflow
velocity scales with mass similarly to the halo circular velocity.
Outflows exhibit a clear bimodal flow pattern, with strong preferential
  alignment along the minor axes pointing orthogonal to galaxy disks.

We find that ISM-scale mass fluxes are not fully converged with numerical
  resolution for $M_{200} < 10^{12} \, \mathrm{M_\odot}$, despite adjusting
  feedback parameters to recalibrate the simulation against the observed galaxy stellar
  mass function (Fig.~\ref{recal_comp}). Convergence is better for outflows
  at the halo scale, and qualitatively our main conclusions hold at eight
  times higher mass resolution for both spatial scales.
  Energy fluxes are well converged at both scales in the recalibrated simulation,
  indicating that energy fluxes are a better indicator of
  the impact of feedback on galaxy assembly than ISM-scale mass fluxes (since the
  galaxy stellar mass functions of the two simulations agree despite discrepant
  ISM-scale mass fluxes).

Comparing to other cosmological hydrodynamical simulations (Fig.~\ref{sim_comparison}), 
we demonstrate that while substantial quantitative differences
are found for gas being driven from the ISM (up to $0.5 \, \mathrm{dex}$), most
simulations show qualitatively similar trends, although for $M_{200} > 10^{12} \, \mathrm{M_\odot}$
this is only the case if AGN feedback is included.
The largest uncertainty in the current picture for outflows comes from the dichotomy between outflow
rates measured at different spatial scales. For example, we show that the \eagle and 
Illustris-TNG simulations present completely different scenarios for gas outflows at 
$50 \, \mathrm{kpc}$ from galaxies versus outflows at $10 \, \mathrm{kpc}$. At $z=2$ and at
a galaxy stellar mass of $10^9 \, \mathrm{M_\odot}$, outflow rates are an order of magnitude
higher at $10 \, \mathrm{kpc}$ than at $50 \, \mathrm{kpc}$ in Illustris-TNG, whereas there is little
difference in flux between these spatial scales in the \eagle simulation.
At high mass ($M_\star \sim 10^{11} \, \mathrm{M_\odot}$), outflows in TNG stay approximately
constant with radius from $10 \, \mathrm{kpc}$ to $50 \, \mathrm{kpc}$, whereas outflow
rates increase with radius in \eagle by nearly an order of magnitude over the same range.  
\eagle therefore presents an ejective (but not ballistic) scenario for galactic winds driven by stellar
feedback, where comparatively few baryons are removed from the ISM but are
driven out to relatively large distances while sweeping up circum-galactic gas. Illustris-TNG instead presents a comparatively more
fountain-like scenario, where more baryons are removed from the ISM
by supernovae but are not driven as far, and so (presumably) can be recycled more efficiently.
\cite{Davies19} have shown that this is reflected in the total baryon content of low-mass ($M_{200} < 10^{12} \, \mathrm{\odot}$)
haloes between the two simulations, with much higher baryon fractions for TNG than for \eagle.
The \fire zoom-in simulations are similar to TNG in the sense that they report
lower outflow rates at the virial radius than at one quarter of the virial radius (although
FIRE agrees well with \eagle at $r=0.25 \, R_\mathrm{vir}$ if the comparison is performed
at fixed halo mass).

% Richard would like to zoom-out on this point to the broader picture from other simulations
% and observaitons

The differences between simulations closely echo the picture encapsulated by semi-analytic
galaxy formation models (Fig.~\ref{sam_comparison}), where acceptable matches to the observed galaxy luminosity function
can be achieved using a very wide range in mass loading factor, with high outflow rates
from the ISM (but not from the halo) being degenerate with high outflow rates
through the halo virial radius (but not from the ISM), and both scenarios also being
degenerate with the timescale for ejected gas to return.
Measurements of the distribution of metals both within and outside galaxies presumably
represent a means to move beyond this impasse, as well as observational estimates of outflow
rates that span a range of spatial scales. 
With some of the clearest differences between simulations
seen for low-mass galaxies ($M_\star \sim 10^9 \, \mathrm{M_\odot}$), the regime where AGN
feedback is not predicted to play an important role, observations that probe metals in
the vicinity of dwarf galaxies may represent a particularly promising avenue to distinguish
between ejective and fountain-like scenarios \cite[e.g.][]{Burchett16,Johnson17}.

\section*{Acknowledgments}

We would like to thank Dr. John Helly for producing and sharing the halo
merger trees that form the backbone of our analysis. We would also like
to thank Ricarda Beckmann for providing measurements of outflow
rates from the Horizon-AGN simulations.

This work used the DiRAC@Durham facility managed by the Institute for
Computational Cosmology on behalf of the STFC DiRAC HPC Facility
(www.dirac.ac.uk). The equipment was funded by BEIS capital funding
via STFC capital grants ST/K00042X/1, ST/P002293/1, ST/R002371/1 and
ST/S002502/1, Durham University and STFC operations grant
ST/R000832/1. DiRAC is part of the National e-Infrastructure.

PM acknowledges the LABEX Lyon Institute of Origins (ANR-10-LABX-0066)
of the Université de Lyon for its financial support within the program
``Investissements d'Avenir'' (ANR-11-IDEX-0007) of the French 
government operated by the National Research Agency (ANR).
This work was supported by Vici grant 639.043.409 from
the Netherlands Organisation for Scientific Research (NWO).
RAC is a Royal Society University Research Fellow.
RGB acknowledges support from the Durham consolidated grant:
ST/P000541/1.

%----------------------------------------------
\bibliographystyle{mn2e}
\bibliography{bibliography}
%---------------------------------------------------------------------

\appendix

\section{Methodology details}
\label{ap_method}

%Here, 

\subsection{Temporal convergence}
\label{ap_converge}

\begin{figure}
\includegraphics[width=20pc]{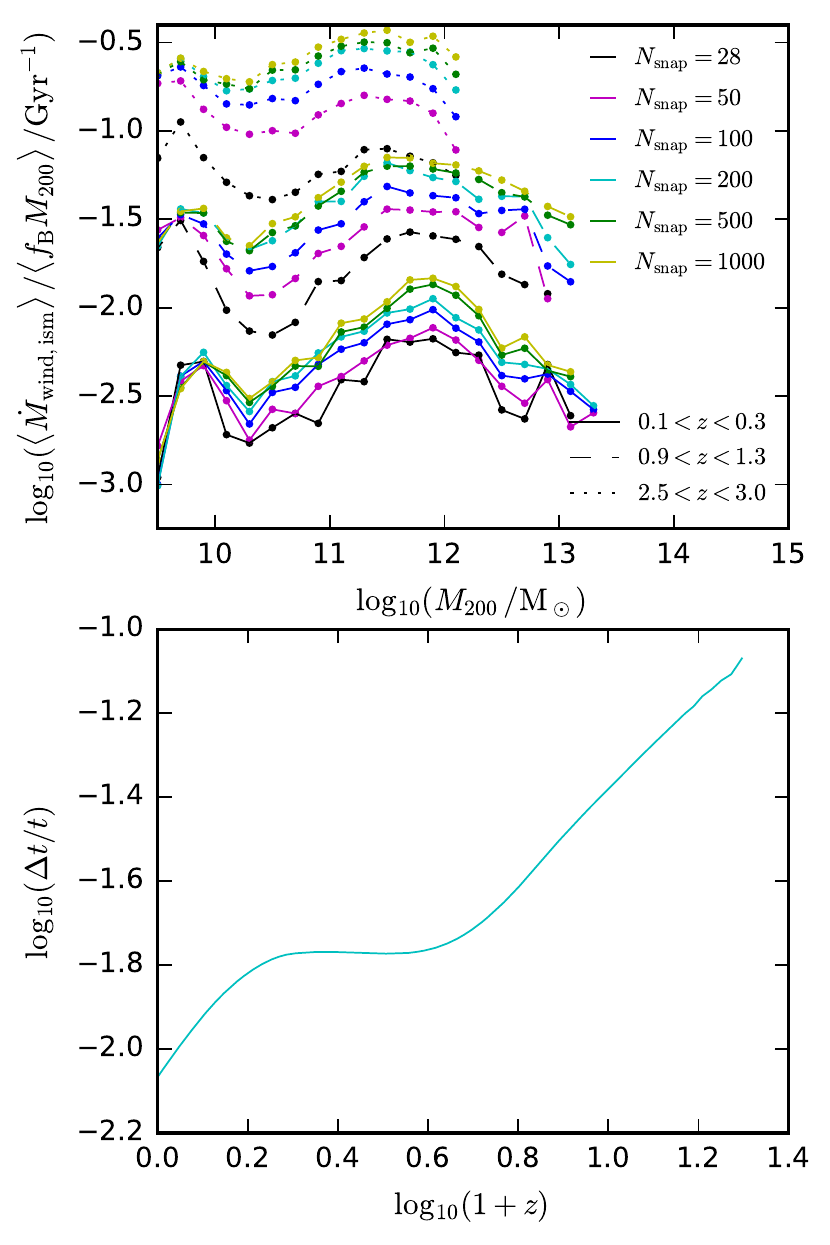}
\caption{Temporal convergence for outflow rates of gas ejected from the ISM.
\textit{Top:} Outflow rate (scaled by halo mass) as a function of halo mass.
Each line colour corresponds to a different number of simulation snapshots used to perform the measurement, as labelled.
Line styles indicate different redshifts.
\textit{Bottom:} the temporal spacing of the 200 simulation outputs used for our fiducial analysis, expressed
as the ratio of the output spacing, $\Delta t$, to the age of the Universe at a given epoch, $t$.
Data are taken from a $25 \, \mathrm{Mpc}$ reference simulation, for which a larger number of processed simulation
snapshots were available.}
\label{convergence_plot}
\end{figure}

An important caveat of Lagrangian flux measurements is that any mass element that crosses the chosen surface
more than once (over the finite time interval adopted) will lead to an underestimate of the flux.
This is particular pertinent for measuring fluxes at the interface to the ISM, where the timescales
for gas to be accreted and then ejected by feedback can be short.
The top panel of Fig.~\ref{convergence_plot} shows the temporal convergence properties for measurements of gas ejected
from the ISM, using a $25^3 \, \mathrm{Mpc}^3$ simulation with a higher frequency of simulation outputs ($1000$ in total, compared
to $200$ for our fiducial simulation).
We find that the measurements start to be reasonably well converged once $200$ simulation outputs are used (cyan lines),
apart from at low redshift, where outflow rates are underestimated by $\approx 0.2 \, \mathrm{dex}$ with respect to the
measurements made using $1000$ snapshots (solid yellow line).
There are only $200$ processed simulation outputs available for the larger $100 \, \mathrm{Mpc}$ reference
simulation, and so we use this set of outputs (and associated merger trees) for our fiducial analysis in this study.
The temporal spacing of these $200$ outputs is shown by the bottom panel of Fig.~\ref{convergence_plot}.

Fig.~\ref{convergence_plot} shows that temporal convergence issues tend to affect outflow rates with a fairly
constant fractional offset as a function of halo mass at a given redshift. The main effect on our results is
that the offset between our measurements of galaxy and halo-scale outflow rates (which are much better converged
due to the longer associated timescales) will increase spuriously at low redshift,
partly explaining the trends seen in Fig.~\ref{mass_loading_ratio}.

\subsection{Lagrangian outflow rates}

We present here a more detailed explanation of how we arrived at the selection criteria 
described in Section~\ref{outflow_description}. 
These criteria were chosen to find a reasonable
balance between completeness and purity, with temporal convergence (as described above) another
consideration. Given that the aim of this study is to measure the flux of gas being evacuated
from the ISM or from the halo, we take genuine outflowing particles to be those
that leave a given component, and then proceed to move a significant distance outwards in radius.

For a number of example galaxies (spanning a wide range in mass and redshift), we compute
the maximum change of radius, $\Delta r_{\mathrm{max}}$, for particles that leave the ISM (or halo virial radius), up until
the time that the particle ceases to be outflowing, or otherwise rejoins the ISM (or halo).
The distribution of $\Delta r_{\mathrm{max}}$ is typically characterised by a peak of particles
that do not move significantly outwards, and then a long extended tail of particles that move
over a wide range of radii. For at least a subset of the total mass and redshift range,
the fraction of particles leaving the ISM (or halo) that do not move a significant distance
is substantial. From detailed inspection of individual particle trajectories, these are often
particles that have recently been accreted onto the ISM but are still in the process
of settling into the disk, and so fluctuate across the ISM boundary a number of times.

We find that adopting cuts in instantaneous velocity or energy \cite[as with the criteria adopted by for example][]{Hopkins12,Christensen16} 
yields relatively poor completeness/purity in terms of the radial distance then traveled by
outflowing particles. At the same time, computing $\Delta r_{\mathrm{max}}$ from the full future radial trajectory of all particles
from a large simulation would be prohibitively expensive. We compromise in this by computing the radial
displacement reached by particles after one quarter of a halo dynamical time
has passed since they left the ISM , which we find to be an excellent proxy for the $\Delta r_{\mathrm{max}}$ computed
using many simulation outputs. This thus motivates our choice of Eqn~\ref{ism_wind_criteria1}
in Section~\ref{outflow_description}.

\subsection{Impact of radial velocity cuts and ISM definition}
\label{vcut_sec}

\begin{figure}
\includegraphics[width=20pc]{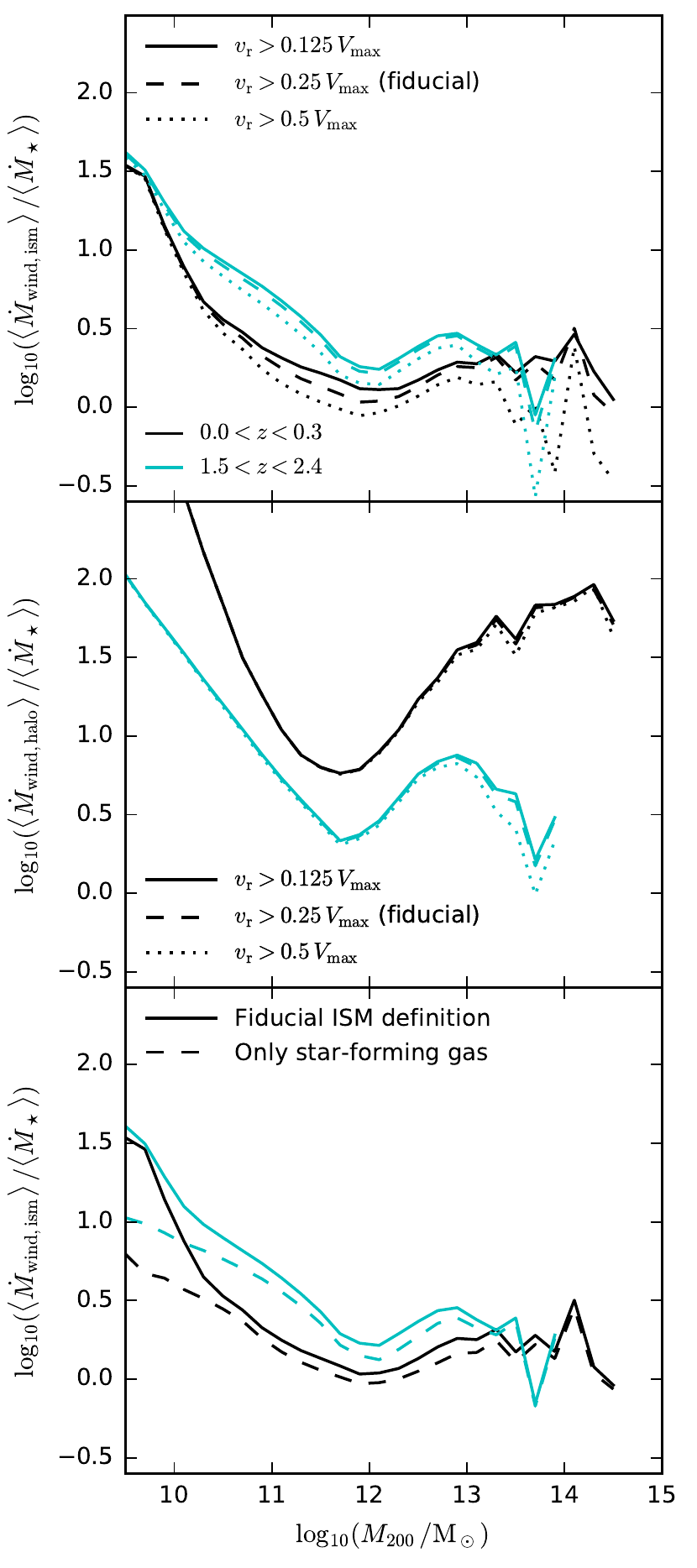}
\caption{The impact of changing the time-averaged radial velocity cut used to select outflowing galaxy-scale wind particles (top/middle), and of changing the ISM definition (bottom).
Adjusting the radial velocity cut has negligible effects on the halo-scale outflows (middle panel).
A factor of two change relative to our fiducial velocity cut of $0.25 \, V_{\mathrm{max}}$ changes the galaxy-scale outflow rates by about $0.1 \, \mathrm{dex}$, although larger differences are seen in high-mass haloes.
The lower panel shows the impact of changing our fiducial ISM definition to a selection of star-forming gas only.
The main effect of including non-star-forming gas in our ISM definition is to enhance the outflow rates in low-mass galaxies
(where metallicities are low and less gas can pass the metallicity-dependent star formation threshold).}
\label{vcut_ism_plot}
\end{figure}

Our results are not highly sensitive to the choice of (time-averaged) radial velocity cut in 
Eqn~\ref{ism_wind_criteria1}. This is demonstrated in Fig.~\ref{vcut_ism_plot}, which shows
that galaxy-scale outflow rates change by small amounts when varying the cut (although there is
a more significant impact on outflow rates in high-mass haloes). Note that removing
the cut completely would have a much larger impact (gas leaving the ISM can often move
inwards over quarter of a halo dynamical time).

The bottom panel of Fig.~\ref{vcut_ism_plot} demonstrates the impact of including/excluding
non-star-forming gas from our ISM criteria. Including this material slightly enhances
the galaxy-scale outflow rates at all mass scales, but the main effect is to substantially
enhance the outflow rates at low halo mass. Given that this is the regime where galaxies
often have not formed a single star particle over the entire redshift interval (meaning results are 
likely not well converged at low mass), the impact on our results is modest.

\subsection{Comparison of Lagrangian and Eulerian fluxes}
\label{method_comparison}

\begin{figure*}
\begin{center}
\includegraphics[width=40pc]{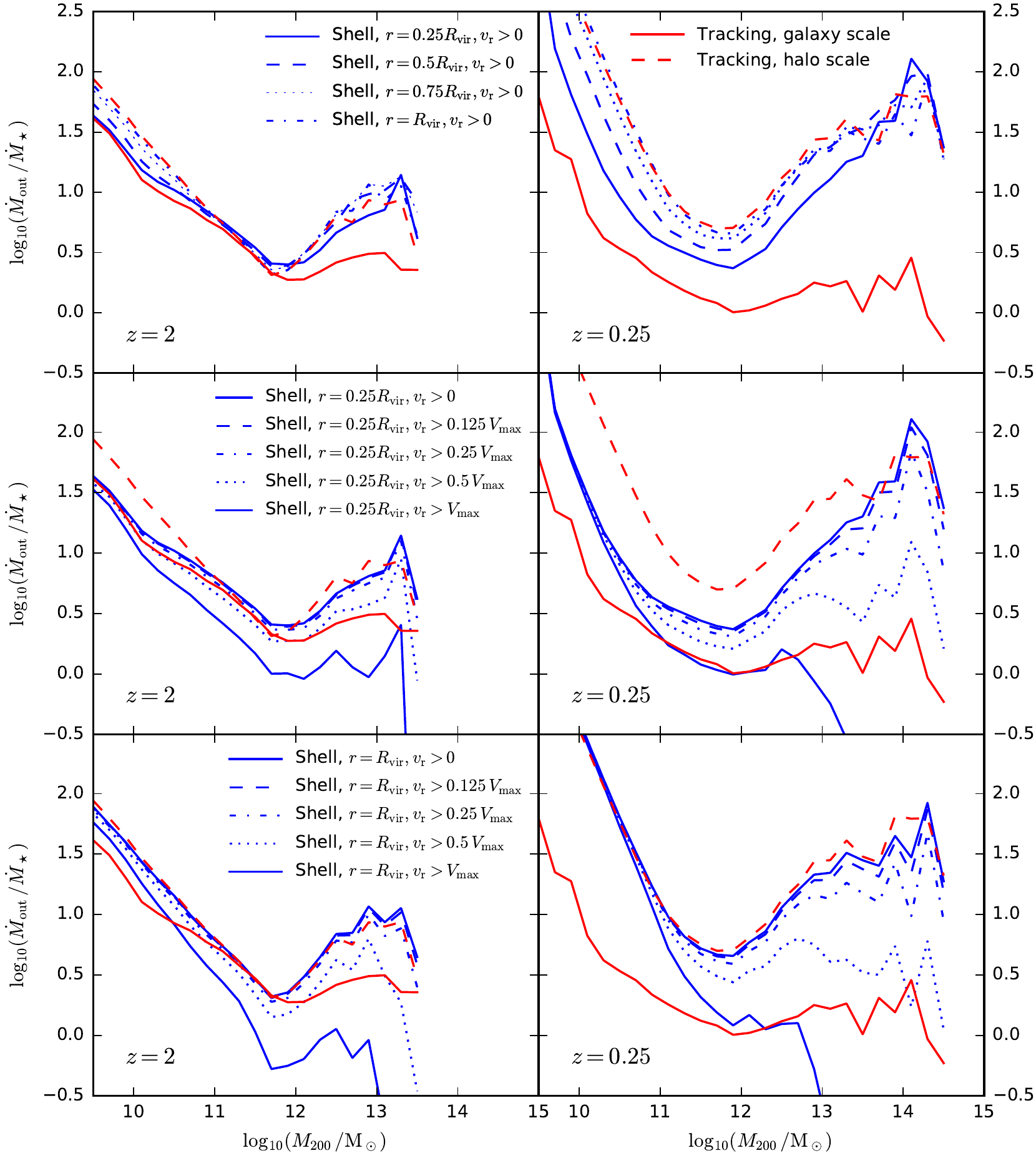}
\caption{
A comparison of our fiducial Lagrangian measurement of wind mass loading factors with simple shell-based measurements, performed at $z=2$ (left) and at $z=0.25$ (right).
Lagrangian outflow rates (red) are shown for gas leaving the ISM (solid), and the halo virial radius (dashed).
Eulerian measurements (blue) are computed from the instantaneous radial momentum, summed over particles within shells of width $0.1 R_{\mathrm{vir}}$, 
including only particles outflowing faster than some minimum (instantaneous) radial velocity.
\textit{Top-row:} a comparison with shells at different radii, selecting all outflowing gas.
\textit{Middle-row:} a comparison with shells at one quarter of the halo virial radius, selecting gas with different radial velocity cuts.
\textit{Bottom-row:} a comparison with shells at the halo virial radius, selecting gas with different radial velocity cuts.
}
\label{method_comp_plot}
\end{center}
\end{figure*}

Fig.~\ref{method_comp_plot} presents a comparison of our Lagrangian measurement of wind mass loading factors with
a simple Eulerian measurement performed by summing the radial momentum of outflowing particles within 
spherical shells. 
For shells placed at the halo virial radius, it is evident that our Lagrangian 
criteria are equivalent to selecting all outflowing gas with $v_{\mathrm{rad}} > 0 \, \mathrm{km s^{-1}}$,
reflecting the looseness of our Lagrangian selection criteria for halo-scale outflows.

Comparing our Lagrangian galaxy-scale outflows to shell-based measurements at one quarter of the halo virial
radius, it is clear that the Lagrangian measurements are always lower. At $z=2$ (top-left), there is some evidence
for entrainment seen in the shell measurements at different radii, with outflow rate increasing with radius
by about $0.3 \, \mathrm{dex}$ in both low and high-mass haloes. %Our Lagrangian ISM-scale outflows are lower
%than at the virial radius partly because of this effect, but also because we impose stricter radial velocity
%thresholds for the ISM-scale outflows (Eqn~\ref{ism_wind_criteria1}). This latter point can be appreciated
%by considering shell measurements with different radial velocity cuts (middle-left). % Note for this it would make more sense to use radial velocity cut at a fraction of vmax
%We note again that the ISM radial velocity cut used for the Lagrangian-scale ISM measurements is not
%arbitrary, and is chosen to accurately select particles that are were ejected from equilibrium in the ISM
%because of feedback. That this makes a significant quantitative difference relative to the shell measurements
%at $0.25 R_{\mathrm{vir}}$ serves to underline the importance of studying the detailed
%trajectories of fluid elements in order to better estimate the flux of gas being evacuated from the ISM by feedback.

At $z=0$ there is an increased entrainment effect seen in the shell measurements, with outflow rates
higher by about $0.5 \, \mathrm{dex}$ at the virial radius when compared to one quarter of the virial radius.
Our Lagrangian measurements for gas leaving the ISM are again lower. 
A lack of temporal convergence at $z=0$ has a small systematic contribution to this effect (see Fig.~\ref{convergence_plot}).

\section{Parametric fits to galaxy-scale and halo-scale mass loading factors}
\label{ap_parafit}

To facilitate comparisons with other studies, we provide parametric fits to the mass loading
factors for both galaxy and halo-scale outflows (as shown in Fig.~\ref{mass_loading}). 
We only fit to data from bins where more than $80\%$ of the galaxies have formed at least one star particle,
integrated over the redshift bins indicated in Fig.~\ref{mass_loading}. 
We find that a reasonable fit to the mass loading factors as a function of halo mass
is given by the five parameter function,

\begin{multline}
\log_{10}\left(\frac{\dot{M}_{\mathrm{out}}}{\dot{M}_\star}\right) = \log_{10}\left(N \, \left[\left(\frac{M_{200}}{M_{\mathrm{1}}}\right)^{\alpha} + \left(\frac{M_{200}}{M_{\mathrm{1}}}\right)^{\beta} \right] \right) \\
+ \delta \, \log_{10}\left(\frac{M_{200}}{M_{\mathrm{1}}}\right) \, \exp\left(-M_{200} / M_{\mathrm{cut}}\right),
\label{para_fit}
\end{multline}

\noindent where for galaxy-scale outflows:

% ISM winds as a function of mhalo
% Anchor at bin 1, valid over 0.06<z<3
\begin{equation}
\begin{split}
\log_{10}(N) = -0.25 + 0.11 \, z \\
\log_{10}(M_1 \, /\mathrm{M_\odot}) = 12.31 \\
\alpha = -0.39 -0.06 \, z \\
\beta = 1.20 \\
\log_{10}(M_{\mathrm{cut}} \, / \mathrm{M_\odot})  = 12.84 \\
\delta = -1.04,
\end{split}
\end{equation}

\noindent where two of the parameters are fit as a linear function of redshift, $z$.
For halo-scale outflows, we find a reasonable fit is given by the same function, but with

% halo winds as a function of mhalo
% Anchor at bin 1, valid over 0.06<z<4
\begin{equation}
\begin{split}
\log_{10}(N) = -0.15 + 0.67 \, a \\
\log_{10}(M_1 \, /\mathrm{M_\odot}) = 11.55 + 0.17 \, z\\
\alpha = -1.19 + 0.18 \, z \\
\beta = 0.74 + 0.26 \, z \\
\log_{10}(M_{\mathrm{cut}} \, / \mathrm{M_\odot}) = 13.46 - 0.32 \, z \\
\delta = -0.27 - 0.45 \, z,
\end{split}
\end{equation}

\noindent where in this case $\log_{10}(N)$ is fit as a function of expansion factor, $a$.

The parameter $\alpha$ sets the low-mass power law slope of the mass loading factor as a function of halo mass
(primarily related to stellar feedback), and $\beta$ sets the power law slope of the upturn at higher masses (primarily
related to AGN feedback). $M_1$ sets the transition halo mass scale between these regimes\footnote{$M_1$ is close but not
exactly equal to the halo mass where the mass loading factor reaches a local minimum value.}, and $N$ sets the overall normalisation.
$\delta$ and $M_{\mathrm{cut}}$ are responsible for the third (flatter or negative) power law slope that becomes evident in group/cluster
mass haloes.
Both of these fits (galaxy and halo scale) provide a reasonable description of the data shown in Fig.~\ref{mass_loading} (within at least $\approx 0.1 \, \mathrm{dex}$) up until $z=4$.

We also provide parametric fits to the galaxy-scale mass loading factor
as a function of galaxy stellar mass and halo maximum circular velocity. We again adopt the five parameter functional
form given by Eqn~\ref{para_fit}, switching the dependent variable from halo mass to maximum circular velocity or stellar mass.

As a function of halo maximum circular velocity, $V_{\mathrm{max}}$, we find

% ISM winds as a function of vmax
% Anchor at bin 1, valid over 0.06<z<3
\begin{equation}
\begin{split}
\log_{10}(N) = -0.31 + 0.14 \, z \\
\log_{10}(M_1 \, / \mathrm{km s^{-1}}) =  2.22 + 0.04 \, z \\
\alpha =  -1.43 - 0.17 \, z \\
\beta =  4.02 \\
M_{\mathrm{cut}} =  161 \, \mathrm{km s^{-1}} \\
\delta = -4.18.
\end{split}
\end{equation}

%logN fit p0,p1 =  -0.3078487666928769 0.13757383046210425 only valid for z> 0.11934602790379656
%logM1 fit p0,p1 =  2.2202782144292668 0.04371790837228774 only valid for z> 0.11934602790379656
%n1 fit p0,p1 =  -1.428253781139405 -0.1738253579630067 only valid for z> 0.11934602790379656
%n2,n3,Mcut,logMcut 4.020940095295047 -4.178158369814651 160.67112116414557 2.2059378242050207
%Fit is valid for range 0.11934602790379656 2.87107463354472

\noindent As a function of galaxy stellar mass (measured within a $30 \, \mathrm{pkpc}$ spherical
aperture), we find

% ISM winds as a function of mstar
% Anchor at bin 2 (the exception), valid over 0.06<z<3
\begin{equation}
\begin{split}
\log_{10}(N) = -0.15 + 0.08 \, z \\
\log_{10}(M_1 \, / \mathrm{M_\odot}) = 10.82 - 0.07 \, z\\
\alpha = -0.22 - 0.06 \, z \\
\beta = 2.10 \\
\log_{10}(M_{\mathrm{cut}} \, / \mathrm{M_\odot}) = 10.83 \\
\delta = -2.27.
\end{split}
\end{equation}

\noindent In both cases, the fits provide a reasonable description of the data up to $z=3$.

%%%%%%%% Data for Vcirc instead of vmax %%%%%%%%%%%%%%

% ISM winds as a function of vcirc
% Anchor at bin 1, valid over 0.06<z<3
%\begin{equation}
%\begin{split}
%\log_{10}(N) = -0.32 + 0.13 \, z \\
%\log_{10}(M_1 \, / \mathrm{km s^{-1}}) = 2.18 + 0.06 \, z \\
%\alpha = -1.46 -0.19 \, z \\
%\beta = 4.47 \\
%M_{\mathrm{cut}} = 255 \, \mathrm{km s^{-1}} \\
%\delta = -5.82.
%\end{split}
%\end{equation}

% Note to self - this is fit to the second reduction, after correcting for the r200_host not being converted from comoving bug (which affected vcirc)
%logN fit p0,p1 =  -0.31516683635732035 0.12749293349118718 only valid for z> 0.11934602790379656
%logM1 fit p0,p1 =  2.180703424016307 0.05583243648393644 only valid for z> 0.11934602790379656
%n1 fit p0,p1 =  -1.458873975556207 -0.19204106677078264 only valid for z> 0.11934602790379656
%n2,n3,Mcut,logMcut 4.467294694223263 -5.815916530385801 254.63061710622364 2.4059106225730895
%Fit is valid for range 0.11934602790379656 2.87107463354472

% LocalWords:  comoving

\section{Comparison to the \fire simulations as a function of halo mass}
\label{ap_fire_comp}

\begin{figure}
\includegraphics[width=20pc]{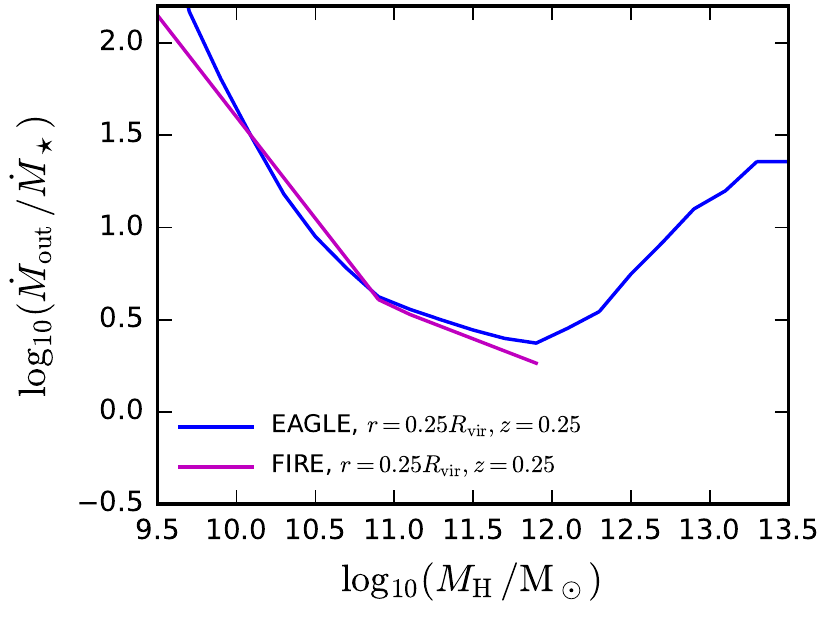}
\caption{A comparison of mass loading factors between \eagle and the \fire simulations, in this case plotted as a function of halo mass.
Mass loading factors are measured in shells with $0.2 < r / R_{\mathrm{vir}} < 0.3$, selecting all outflowing gas.
The magenta line shows the best-fit relation from \protect \cite{Muratov15} plotted at $z=0.25$.
The blue line shows the average mass loading factor from \eagle, also plotted at $z=0.25$.
Plotted as a function of halo mass, \eagle and \fire are in excellent agreement over the common mass range at $r=0.25 R_{\mathrm{vir}}$.}
\label{fire_mhalo_comp}
\end{figure}

Further to the comparison between simulations shown in Fig.~\ref{sim_comparison} of Section~\ref{sim_comp_sec}, Fig.~\ref{fire_mhalo_comp}
shows a comparison between the mass loading factors in \fire and \eagle at low redshift, measured at $r=0.25 \, R_{\mathrm{vir}}$,
and plotted in this case as a function of halo mass, rather than as a function of stellar mass.
Compared at a given halo mass, \eagle and the best-fit relation from \fire are in remarkably good agreement over
the common mass range. The level of agreement is significantly better than when the simulations
are compared as a function of galaxy stellar mass (as shown in Fig.~\ref{sim_comparison}), where
the best-fit mass loading factor relation in \fire is about $0.3 \, \mathrm{dex}$ higher than the
average from \eagle at $M_\star \sim 10^{9} \, \mathrm{M_\odot}$. This implies there is a systematic
difference in the median stellar mass versus halo mass relation between the two sets of simulations.

Individual galaxies in \fire at $M_{\mathrm{H}} \sim 10^{12} \, \mathrm{M_\odot}$
fall below the plotted best-fit relation at low redshift, and are observed to be relatively quiescent
in terms of outflow activity,
with residual outflowing flux attributed to non-feedback sources \cite[][]{Muratov15}.
This is the mass scale where AGN feedback (which is not implemented in the \fire simulations)
starts to play a significant role in \eagle, causing the upturn of the mass loading factor
at higher halo masses.

% Appendix to give more detailed justification for wind selection criteria

%\section{Appendicies}

%\begin{figure*}
%\begin{center}
%\includegraphics[width=40pc]{Figures/mass_loading_mstar_vmax_dep_L0025N0376_REF_50_snip.pdf}
%\caption{
%Mean outflow rates from the ISM (top panels) and haloes (bottom panels) for central galaxies, plotted as a function of stellar mass (left panels) and maximum circular velocity (right panels). 
%Stellar mass used as the mass variable is (for now) not including an aperture correction.
%At present, data are taken from the $25 \, \mathrm{Mpc}$ reference run, using trees with 50 snipshots.
%}
%\label{mass_loading_mstar_vmax}
%\end{center}
%\end{figure*}

%Fig.~\ref{mass_loading_mstar_vmax}. Outflow rates as a function of stellar mass and maximum circular velocity.

\label{lastpage}
\end{document}